\documentclass{aastex631}
\usepackage{CJK}
\usepackage{graphicx}
\usepackage{enumerate}
\usepackage{amssymb, amsmath}
\usepackage{natbib}
\usepackage{color}
\usepackage{ulem}
\usepackage{dblfnote}
\usepackage{appendix}
\usepackage{hyperref}
\usepackage[stable]{footmisc}
\usepackage{comment}
\usepackage{footnote}

\usepackage[nodayofweek]{datetime}
\newdateformat{monthyearday}{%
  \THEYEAR\ \monthname[\THEMONTH] \twodigit{\THEDAY}}

\newcommand{\mj}{\ensuremath{\,M_{\rm Jup}}}

\newcommand{\mum}{$\mu$m}
\newcommand{\apsa}{Fomalhaut }
\newcommand{\degree}{$^\circ$}
\newcommand{\sone}{\textit{S1}}
\newcommand{\stwo}{\textit{S2}}
\usepackage{soul}
\usepackage[caption=false]{subfig}
\usepackage[
    starfontserif % comment for sans glyphs
    ]{starfont}
\usepackage{amsmath}

\DeclareSymbolFont{starfontsym}{OT1}{sts}{m}{n}
\DeclareMathSymbol{\mathTerra}{\mathord}{starfontsym}{76}

\begin{document}
\received{July 21 2023}
\revised{October 9 2023}
%\accepted{\monthyearday\today}
%\published{}
\submitjournal{AAS Journals}
\shorttitle{JWST/NIRCam Observations of Fomalhaut}
%PUT BACK
\shortauthors{Ygouf et al.}

%\newcommand{\geoff}[1]{\textcolor{purple}{#1}}
%\newcommand{\jorge}[1]{\textcolor{green}{#1}}
%\newcommand{\marie}[1]{\textcolor{blue}{#1}}
%\newcommand{\chas}[1]{\textcolor{orange}{#1}}
%\newcommand{\alex}[1]{\textcolor{brown}{#1}}
%\newcommand{\tbd}{\textcolor{purple}{TBD}}

%\begin{CJK*}{UTF8}{gbsn}

\title{Searching for Planets Orbiting Fomalhaut with JWST/NIRCam}

%% Note that the corresponding author command and emails has to come
%% before everything else. Also place all the emails in the \email
%% command instead of using multiple \email calls.
%PUT BACK
\correspondingauthor{Marie Ygouf, marie.ygouf@jpl.nasa.gov}

\author[0000-0001-7591-2731]{Marie Ygouf}
\affiliation{Jet propulsion Laboratory, California Institute of Technology, Pasadena, CA 91109}

\author[0000-0002-5627-5471]{Charles Beichman}
\affiliation{NASA Exoplanet Science Institute, IPAC, Pasadena, CA 91125}
\affiliation{Jet propulsion Laboratory, California Institute of Technology, Pasadena, CA 91109}

\author[0000-0002-3414-784X]{Jorge Llop-Sayson}
\affiliation{Department of Astronomy, California Institute of Technology, 1200 E. California Boulevard, Pasadena, CA, 91125, USA}

\author[0000-0001-5966-837X]{Geoffrey Bryden}
\affiliation{Jet propulsion Laboratory, California Institute of Technology, Pasadena, CA 91109}

\author[0000-0002-0834-6140]{Jarron Leisenring}
\affiliation{Steward Observatory, University of Arizona, Tucson, AZ, 85721}

\author[0000-0001-8612-3236]{Andras Gaspar}
\affiliation{Steward Observatory, University of Arizona, Tucson, AZ, 85721}

\author{John Krist}
\affiliation{Jet propulsion Laboratory, California Institute of Technology, Pasadena, CA 91109}

\author[0000-0002-7893-6170]{Marcia Rieke}
\affiliation{Steward Observatory, University of Arizona, Tucson, AZ, 85721}

\author[0000-0003-2303-6519]{George Rieke}
\affiliation{Steward Observatory, University of Arizona, Tucson, AZ, 85721}

\author[0000-0002-9977-8255]{Schuyler Wolff}
\affiliation{Steward Observatory, University of Arizona, Tucson, AZ, 85721}

\author[0000-0002-6730-5410]{Thomas Roellig}
\affiliation{NASA Ames Research Center, Mountain View, CA, 94035}

\author[0000-0002-3532-5580]{Kate Su}
\affiliation{Steward Observatory, University of Arizona, Tucson, AZ, 85721}

\author[0000-0003-4565-8239]{Kevin Hainline}
\affiliation{Steward Observatory, University of Arizona, Tucson, AZ, 85721}

\author[0000-0003-0786-2140]{Klaus Hodapp}
\affiliation{University of Hawaii, Hilo, HI, 96720}

\author[0000-0002-8963-8056]{Thomas Greene}
\affiliation{NASA Ames Research Center, Mountain View, CA, 94035}

\author[0000-0003-1227-3084]{Michael Meyer}
\affiliation{University Of Michigan, Madison, MI}

\author{Doug Kelly}
\affiliation{Steward Observatory, University of Arizona, Tucson, AZ, 85721}

\author{Karl Misselt}
\affiliation{Steward Observatory, University of Arizona, Tucson, AZ, 85721}

\author[0000-0003-2434-5225]{John Stansberry}
\affiliation{Space Telescope Science Institute, 3700 San Martin Drive, Baltimore, MD 21218, USA}

\author[0000-0003-4850-9589]{Martha Boyer}
\affiliation{Space Telescope Science Institute, 3700 San Martin Drive, Baltimore, MD 21218, USA}

\author[0000-0002-6773-459X]{Doug Johnstone}
\affiliation{NRC Herzberg Astronomy and Astrophysics, 5071 West Saanich Rd, Victoria, BC, V9E 2E7, Canada}
\affiliation{Department of Physics and Astronomy, University of Victoria, Victoria, BC, V8P 5C2, Canada}

\author[0000-0001-9886-6934]{Scott Horner}
\affiliation{NASA Ames Research Center, Mountain View, CA, 94035}

\author[0000-0002-7162-8036]{Alexandra Greenbaum}
\affiliation{IPAC, California Institute of Technology, Pasadena, CA 91125}

\begin{abstract}

We report observations with the JWST/NIRCam coronagraph of the Fomalhaut ($\alpha$ PsA) system.
This nearby A star hosts a complex debris disk system discovered by the IRAS satellite.  Observations in F444W
and F356W filters using the round 430R mask achieve a 
contrast ratio of $\sim4 \times 10^{-7}$ at 1\arcsec\ and $\sim4 \times 10^{-8}$ outside of 3\arcsec. These observations reach a sensitivity limit $<$1\mj\ across most of the disk region. Consistent with the hypothesis that Fomalhaut b is not a massive planet but is a dust cloud from a planetesimal collision, we do not detect it in either F356W or F444W (the latter band where a Jovian-sized planet should be bright). 
We have reliably detected 10 sources  in and around Fomalhaut
and its debris disk, all but one of which are coincident with Keck or HST sources seen in earlier coronagraphic imaging; we show them to be background objects, %\textbf{including the ``Great Dust Cloud" identified in MIRI data.}
including the ``Great Dust Cloud" identified in MIRI data. However, one of the  objects, located at the edge of the inner dust disk seen in the MIRI images, has no obvious counterpart in imaging at earlier epochs and has a relatively red [F356W]-[F444W]$>$0.7  mag (Vega) color. Whether this object is a background galaxy, brown dwarf, or a Jovian mass planet in the Fomalhaut system will be determined by an approved Cycle 2 follow-up program. Finally, we set upper limits to any scattered light from the outer ring, placing a weak limit on the dust albedo at F356W and F444W.

\end{abstract}

%\tableofcontents

%\newpage

\section{Introduction}

At a distance of only 7.7 pc, the young ($\sim$500 Myr; \cite{Mamajek2012}, \cite{nielsen2019}) and bright (V=1.16 mag) A3V star,  \apsa ($\alpha$ PsA, HR~8728) was one of the original debris disk systems discovered by the IRAS satellite through the strong infrared excess at wavelengths longward of  12 \mum\ \citep{Aumann1985,Gillett1986}. The debris disk phenomenon was soon recognized as a remnant of the planet formation process \citep{Wyatt2008}. When the natal cloud dissipates, in addition to young planets there can be zones where planets did not form and that are occupied by reservoirs of small solid bodies that collide and grind each other down into small dust grains a few 10s to 100s of microns in size. These grains are  heated by the star to emit prominently at mid- and far-infrared wavelengths. The grains are ultimately  lost by  the system via a number of mechanisms,  slowly depleting the reservoir of solid material \citep{Wyatt2008}. 

Debris disks are common among main sequence K- through A-stars
\citep{Su2006, Carpenter2009, Eiroa2013, Thureau2014}.
At 24 $\mu$m, the phenomenon persists for $\sim$500 Myr, after which it decays to much lower levels  \citep{Rieke2005,gaspar2013}.  The far-infrared emission persists for much longer, to $\sim$ 4 Gyr \citep{sierchio2014}. For  main sequence (FGK but excluding late K) stars the fractional incidence of detectable cold debris disks is independent of spectral type, roughly  20\% at current sensitivity levels \citep{sierchio2014}. % The relative excess can be parameterized by the ratio of integrated dust luminosity to stellar luminosity, $L_{dust}/L_\star \simeq 10^{-6}$ to $10^{-3}$.

Only a few stars are close enough and have bright enough debris systems so that their disks can be resolved at  wavelengths from the submillimeter to the visible, typically revealing material distributed in one or more narrow (few AU) rings separated by some 10s of AU. \apsa is prominent among these. Imaging with the Hubble Space Telescope \citep{Kalas2005,Gaspar2020} combined with  infrared  observations with Spitzer, Herschel \citep{Karl2004,Acke2012,Su2013}, JWST  \citep{Gaspar2023} and submillimeter observations from the James Clerk Maxwell Telescope \citep{Holland2003} and ALMA \citep{boley2012, Su2016,White2017, MacGregor2017} have led to detailed understanding of the distribution of dust orbiting Fomalhaut. As shown by \citet{Gaspar2023}, there is an outer ``Kuiper Belt" ring at 140 AU with T$\sim$ 50K, a second interior ring plus a broad distribution of heated dust extending inward toward the region thermally equivalent to the asteroid belt in the Solar System. \apsa also possesses a source of hot dust emission (T$\sim$1700K) seen via interferometry at $\sim$2 \mum\ that indicates the presence of hot dusty grains located within 6 AU from Fomalhaut~\citep{Absil2009}. 
%Absil et al.: "We argue that the thermal emission from hot dusty grains located within 6 AU from Fomalhaut is the most plausible explanation for the detected excess."

\begin{deluxetable*}{llll}[b!]
\tabletypesize{\scriptsize}
\tablewidth{0pt}
\tablecaption{Properties  of the Host Star \apsa \label{tab:star}
}
\tablehead{
\colhead{Property} & \colhead{Value}& \colhead{Units}& \colhead{Comments} }
\startdata
Spectral Type & A3 Va & &\citet{Gray2006,Mamajek2012}\\
T$_{\rm eff}$ &8590$\pm$73 & K &\citet{Mamajek2012}\\
Mass &1.92$\pm$0.02 & M$_\odot$&\citet{Mamajek2012}\\
Luminosity & 16.6$\pm$0.5& L$_\odot$&\citet{Mamajek2012}\\
Age$^a$ &440 &Myr & \citet{Mamajek2012}\\
$[$Fe/H$]$ &0.05$\pm$0.04& dex &\citet{Gaspar2016}\\
%log(g)&4.32$\pm$0.06&cgs &\citet{Maire2020}\\	 
R.A.\ (Eq 2000; Ep 2000)&22$^h$57$^m$39.0$^s$ & & \citet{Hipparcos2007} \\
Dec.\ (Eq 2000; Ep 2000)&$-$29$^o$37$^\prime$20.05\arcsec & & \citet{Hipparcos2007} \\
% DR3 parallax is 31.2191 +- 0.0240
Distance & 7.70$\pm$0.03 & pc & \citet{Hipparcos2007}\\
Proper Motion ($\mu_\alpha,\mu_\delta$)&(328.95,$-$164.67) &mas/yr &\citet{Hipparcos2007}\\
V & 1.155$\pm$0.005 & mag&\citet{{mermilliod1994}}\\
J  & 1.054$\pm$0.02& mag &\citet{carter1990}\\
H  & 1.010$\pm$0.02& mag &\citet{carter1990}\\
K  & 0.999$\pm$0.02& mag &\citet{carter1990}\\
L  & 0.975$\pm$0.05& mag &\citet{carter1990}\\
%\hline
%\hline
%F250M flux & 1.0$\pm$0.02 & Jy & BOSZ stellar model \\
\enddata
\tablecomments{$^a$\citet{nielsen2019} cite a slightly older age  of 550$\pm$70 Myr. For easier comparison with other analyses we have  adopted the 440 MYr age of \citet{Mamajek2012}.}
\end{deluxetable*}

Multiple HST visits led to the detection of a co-moving source located interior to the  ring --- a candidate planet denoted \apsa b \citep{Kalas2008}. Subsequent observations have called its planetary nature into question: the lack of infrared emission proportional to the visible brightness \citep{Currie2013};   a highly elliptical orbit  projected to intersect the ring  in a manner inconsistent with a stable planetary object \citep{Beust2014}; no evidence for \apsa\  b was found in multiple epochs of Spitzer imaging \citep{Marengo2009, janson2015};  and subsequent  HST  observations showing  the object expanding in size and decreasing in brightness \citep{Gaspar2020}. The most likely explanation for the nature of \apsa b is  a slowly dissipating, expanding remnant of a  collision of two planetesimals  \citep{Lawler2015,Gaspar2020}. 

No planets have as yet been detected within the Fomalhaut system. However, there are strong indications of unseen planets. Two have been invoked to shepherd the outer debris ring \citep[e.g.,][]{boley2012}, while a third one is likely responsible for the configuration of the inner debris ring \citep{Gaspar2023}. Detection limits from observations of this system indicate that those planets are less massive than 3\mj\ \citep{Currie2013, Kenworthy2013,janson2015}. Theoretical estimates for the masses of the shepherding planets are smaller than these limits, $\leq$ 1 $M_{Saturn}$.

In summary, \apsa may host a complex planetary system, as reflected in its debris rings, the planetesimal collision that created \apsa b, and the indications of unseen planets. Thus,  it was with two goals in mind that we made \apsa\ the target of a Guaranteed Time program with the James Webb Space Telescope (PID\#1193) employing both the NIRCam and MIRI instruments: 1) search for planets within the Fomalhaut system at 3-5 \mum, including a definitive measurement of \apsa b; and 2) characterize the  dust structures from 3 to 25 \mum. This paper concentrates on the search for planets using NIRCam~\citep{rieke23}.  \citet{Gaspar2023} focus on the properties of the debris disk using MIRI~\citep{wright2023}. 

\section{NIRCam Observations}

On 2022-10-22  we observed \apsa at two roll angles in F444W and F356W filters using the round F430R mask with an inner working angle (IWA) of $\sim$ 0.85\arcsec\ \citep{Krist2010}. The exposure time at F444W was chosen to search for planets down to $<1$ \mj\ mass at 2\arcsec\ and beyond ($\sim$30 AU) assuming a 5 nm wavefront drift and using representative  models of the emission from young gas giants \citep{Marley2021}.  The F356W observations were  made  to a depth adequate to identify and reject   background stars or extragalactic objects based on their [F356W]--[F444W] color. The  integration time in F356W was about a factor of two lower than at F444W to take advantage of the rising Spectral Energy Distribution (SED) of stars to shorter wavelengths.  

Those observations were obtained in two Fields of View (FoV): with the SUB320 sub-array (20\arcsec $\times$ 20\arcsec ) selected to avoid saturation to search for companions as close to the star as possible, and in full array (2.2\arcmin $\times$ 2.2\arcmin) mode to search for companions up to and beyond the outer ring, located at 140 AU. %$>$140 AU ($>$18\arcsec). 
The large FoV  observations also have the potential to detect scattered emission from the  outer ring, i.e.\ the near-IR counterpart of the ring seen by HST, depending on the properties of the dust grains.

We adopted the star  $\delta$  Aqr (HR8709), an A3Vp star with K$_s$=3.06 mag as a Point Spread Function (PSF)  reference. The star is at a separation of 13.8\degree\ on the sky, but for the observing date in question, 2022-Oct-22, the change in solar illumination angle between the two stars is $\sim$7.1\degree\ which helps to minimize the thermal drift in the telescope \citep{Perrin2018}.  We chose a $\sim$$2\times$ longer exposure on the  reference star to  obtain a closer match of SNR in PSF for both targets. Table~\ref{tab:exposures} describes the observing parameters for the deep imaging part of the NIRCam program. To account for the uncertainty in positioning the star in the center of the coronagraphic mask, we used the small grid dithering (SGD) strategy~\citep{Lajoie2016} with the 5-POINT small ($\sim$10–20 mas) dither pattern on the reference star to increase the diversity in the PSF for post-processing and thus to increase the contrast gain at close separation. We maintained a similar SNR per frame for both targets by carefully choosing the detector readout modes, number of groups and integrations.

\begin{deluxetable*}{lllllll}
\tabletypesize{\scriptsize}
\tablewidth{0pt}
\tablecaption{NIRCam Deep Imaging Observing Parameters (PID:\#1193)\label{tab:exposures}
}
\tablehead{
\colhead{Target} & \colhead{Filter}& \colhead{Readout} & \colhead{Groups/Int} & \colhead{Ints/Exp}& \colhead{Dithers} & \colhead{Total Time (sec)}}
\startdata
\apsa (Roll 1; Obs\#14) &F356W/Mask 430R  &FULL/RAPID  &2&13&1 &408 \\
\apsa (Roll 1; Obs\#14) &F444W/Mask 430R  &FULL/RAPID  &2&24&1 &762 \\
\apsa (Roll 1; Obs\#15) &F356W/Mask 430R  &SUB320/RAPID  &3&105&1 & 451 \\
\apsa (Roll 1; Obs\#15) &F444W/Mask 430R  &SUB320/BRIGHT2  &2&116&1 & 901 \\ \hline
\apsa (Roll 2; Obs\#16) &F356W/Mask 430R  &SUB320/RAPID  &3&105&1 & 451\\
\apsa (Roll 2; Obs\#16) &F444W/Mask 430R  &SUB320/BRIGHT2  &2&116&1 & 901 \\
\apsa (Roll 2; Obs\#17) &F356W/Mask 430R  &FULL/RAPID  &2&13&1 &408 \\
\apsa (Roll 2; Obs\#17) &F444W/Mask 430R  &FULL/RAPID  &2&24&1 &762 \\
 \hline
HR8709 (Obs\#18) &F356W/Mask 430R&SUB320/RAPID  &9&17&5 & 910 \\
HR8709 (Obs\#18) &F444W/Mask 430R&SUB320/BRIGHT2  &9&18&5 & 1830 \\
HR8709 (Obs\#19) &F356W/Mask 430R&FULL/RAPID&2&5&5 &751\\
HR8709 (Obs\#19) &F444W/Mask 430R &FULL/RAPID&3&7&5 &1449 \\
\enddata
\tablecomments{The NIRCam program was executed on 2022-20-22 (2022.808)}
\end{deluxetable*}

\section{Data Reduction and Post-Processing}

\subsection{Pipeline Processing}

\begin{figure*}[b!]
\centering
\includegraphics[clip,width=0.36\textwidth]{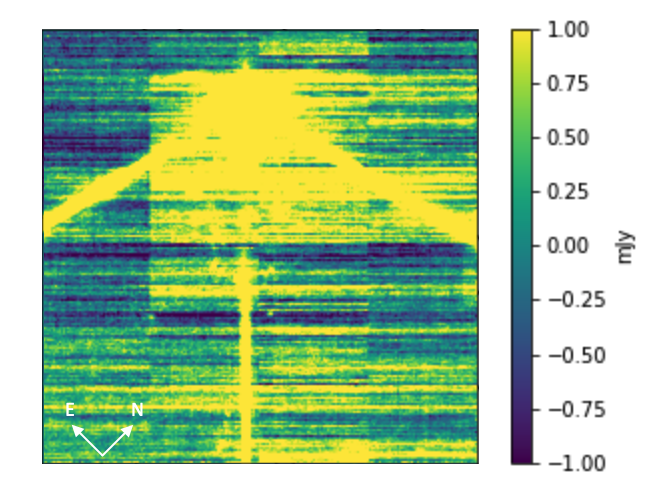}
\includegraphics[clip,width=0.35\textwidth]{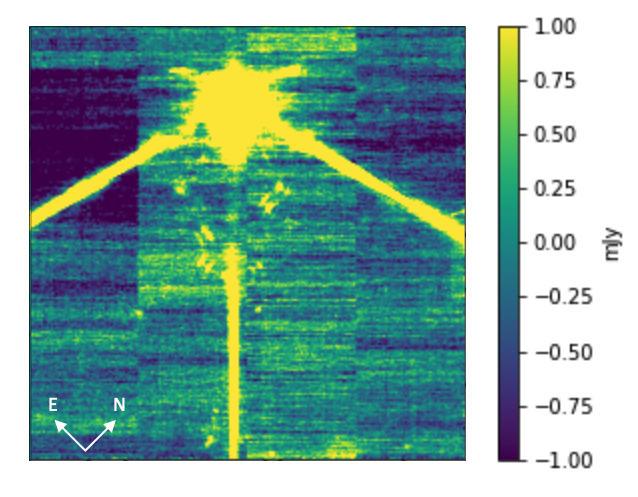}
\caption{Single F356W full frame images ($2.2'\times 2.2'$): raw (left) and corrected from horizontal stripes (right).
\label{fig:raw_data}}
\end{figure*}

\begin{figure*}[b!]
\centering
\includegraphics[clip,width=0.35\textwidth]{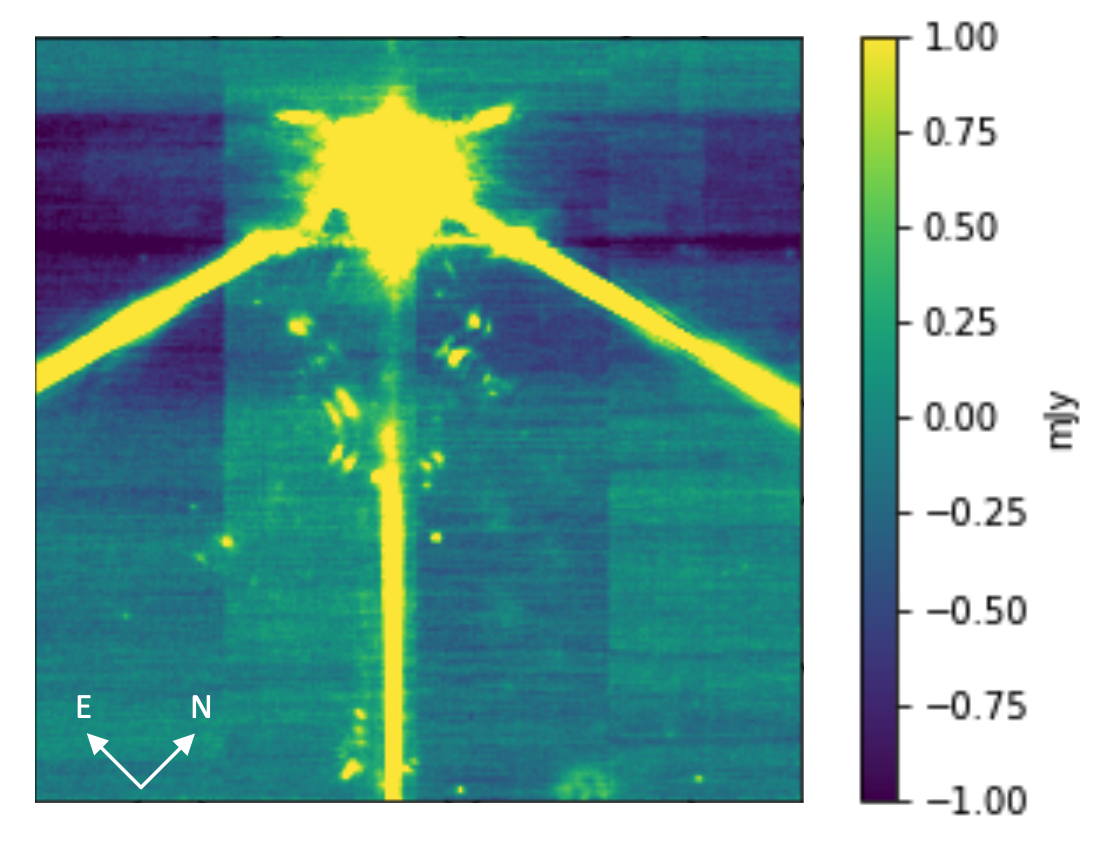}
\includegraphics[clip,width=0.35\textwidth]{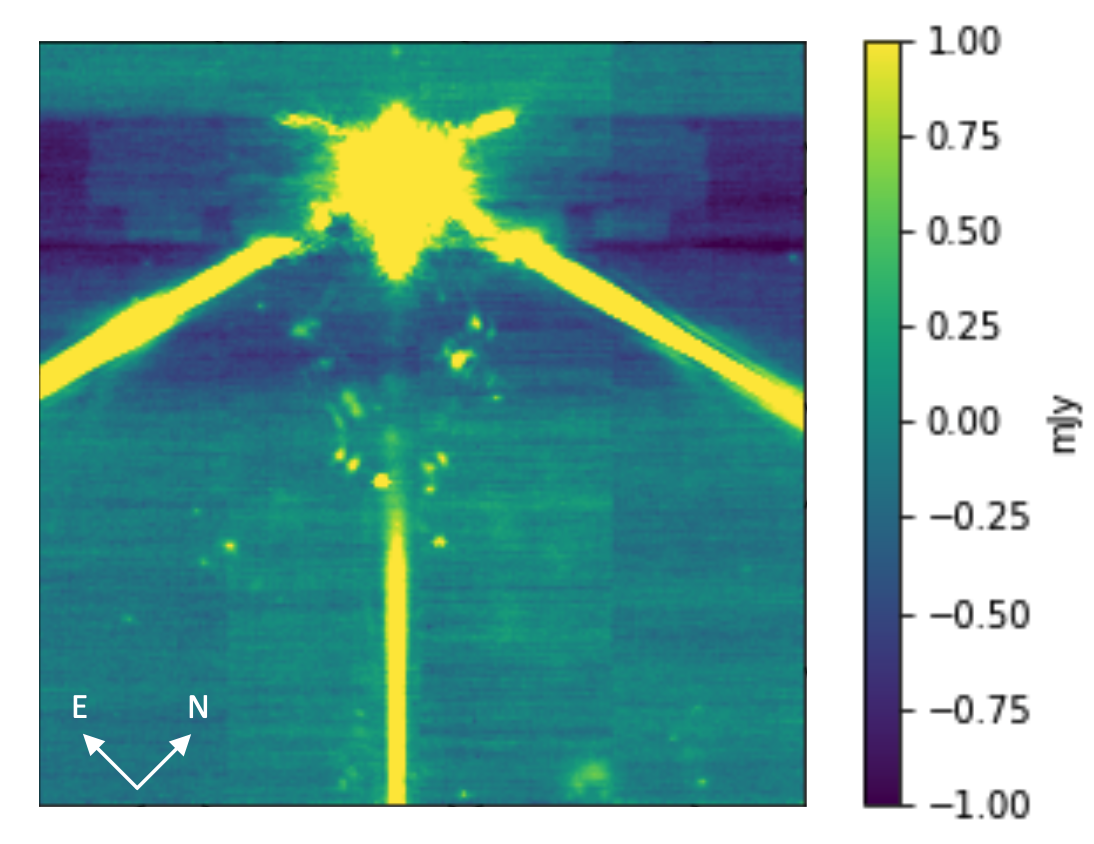}
% \vskip -0.4in
\caption{Coadded full frame images ($2.2'\times 2.2'$) for F356W (left) and F444W (right) before post-processing to remove residual starlight.
\label{fig:coadd_data}}
\end{figure*}

The full set of images (summarized in Table~\ref{tab:exposures})
was processed using the {\tt JWST} pipeline (version 
\rm{2022\_4a}, calibration version
\rm{jwst\_1019.pmap}). The dataset can be obtained at:\dataset[https://doi.org/10.17909/kckm-n422]{https://doi.org/10.17909/kckm-n422}. We started from the uncalibrated stage-1 data products as produced by the standard pipeline \citep{jwst2022} with some modifications of the subsequent steps.
Specifically, 1) we did not include dark current corrections, which are not well characterized for subarray observations;
2) we performed a modified version of the ramp fitting, as implemented within the {\tt SpaceKLIP} package \citep{Kammerer2022} to significantly improve the noise floor in the subarray images; and 3) to reduce saturation effects in the full array images, we allowed fluxes to be measured for ramps that only have a single group before saturation.

\subsection{Bad Pixel Rejection and Horizontal Striping Removal}

We performed additional steps to reject bad pixels and remove horizontal striping resulting from 1/f noise. The standard pipeline removes pixels adjacent to saturation-flagged pixels, which can result in an overly aggressive removal of good data. Since the images do not have any evidence of charge spillage, we utilized less conservative flagging ({\tt n\_pix\_grow\_sat} set to 0, rather than 1).
For identification of truly bad pixels, we used the pipeline {\tt DQ} flags: any pixels flagged as {\tt DO\_NOT\_USE}, e.g.\ dead pixels, those without a linearity correction, etc.,  were set to {\tt NaN}. 5-$\sigma$ outliers -- temporally within sub-exposures or spatially within a 5x5 box -- were also rejected.

\begin{figure*}[t!]
\centering
\includegraphics[clip,width=0.45\textwidth]{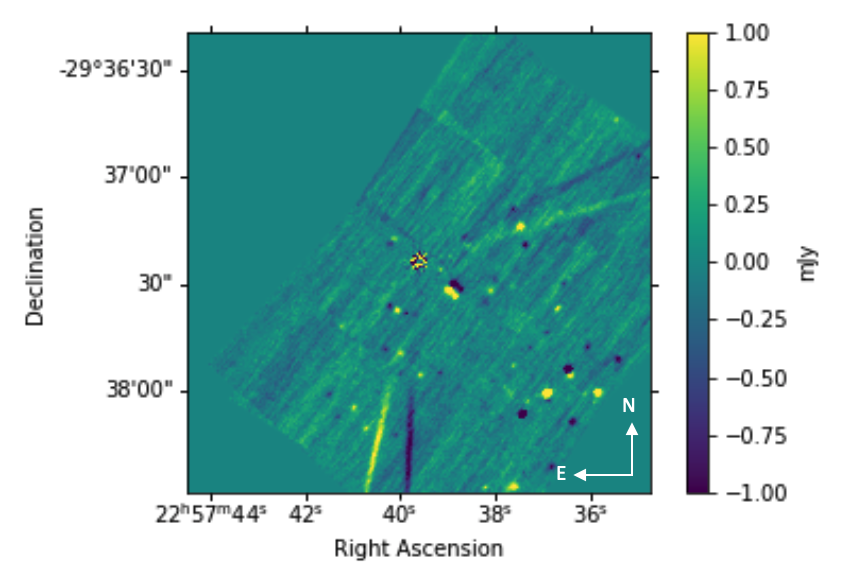}
\includegraphics[clip,width=0.45\textwidth]{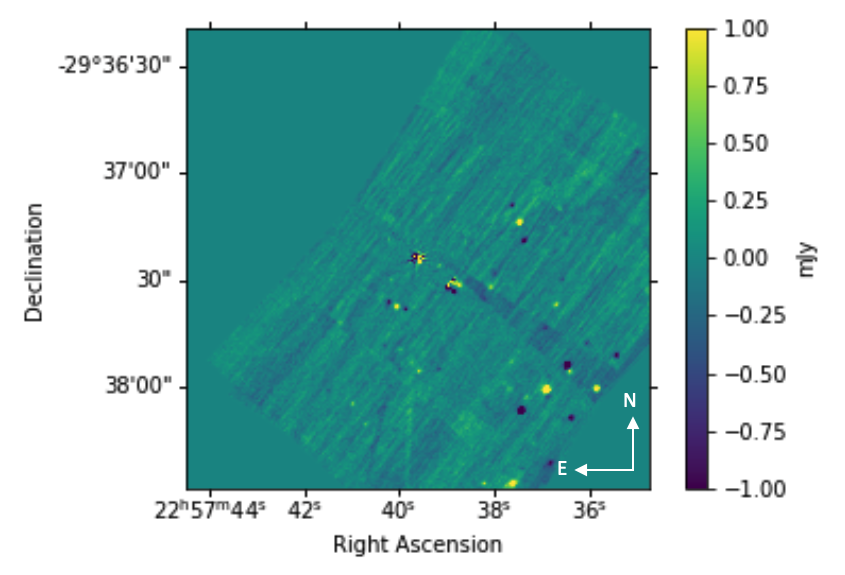}
% \vskip -0.4in
\caption{Classical ADI reductions for the full-frame datasets for F356W (left) and F444W (right). The negative-positive-negative images surrounding each source are due to the subtraction of the two rolls as discussed in the text.
\label{fig:reduced_images_cADI}}
\end{figure*}

\begin{figure*}[t!]%
     \centering
     % \subfloat[]{\centering\includegraphics[width=0.9\textwidth]{fomalhaut_pca_allFilters.png}
     \subfloat[]{\centering\includegraphics[width=16cm]{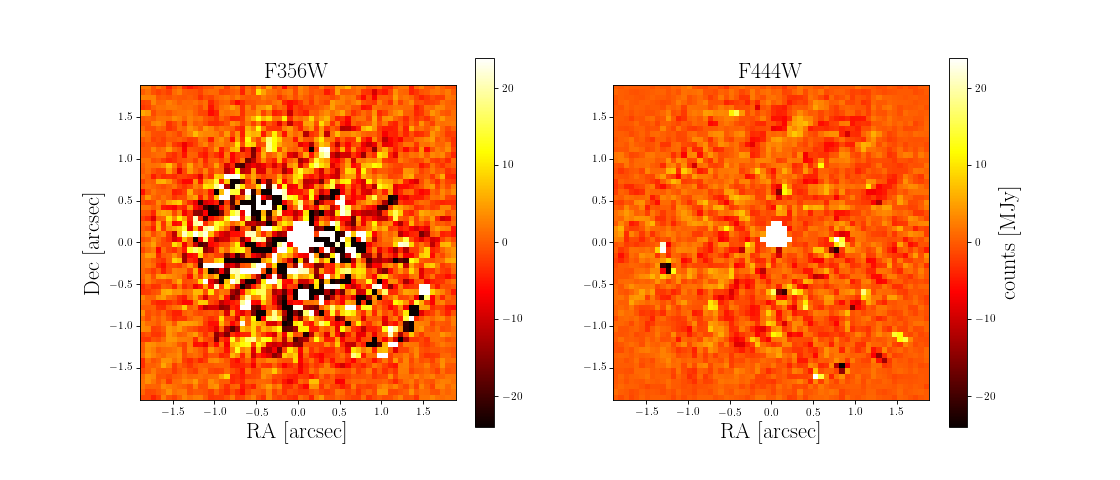}
     \label{fig:top_pca}}\\
     % \subfloat[]{\centering\includegraphics[width=0.9\textwidth]{fomalhaut_fmmf_allFilters.png}
    \subfloat[]{\centering\includegraphics[width=16cm]{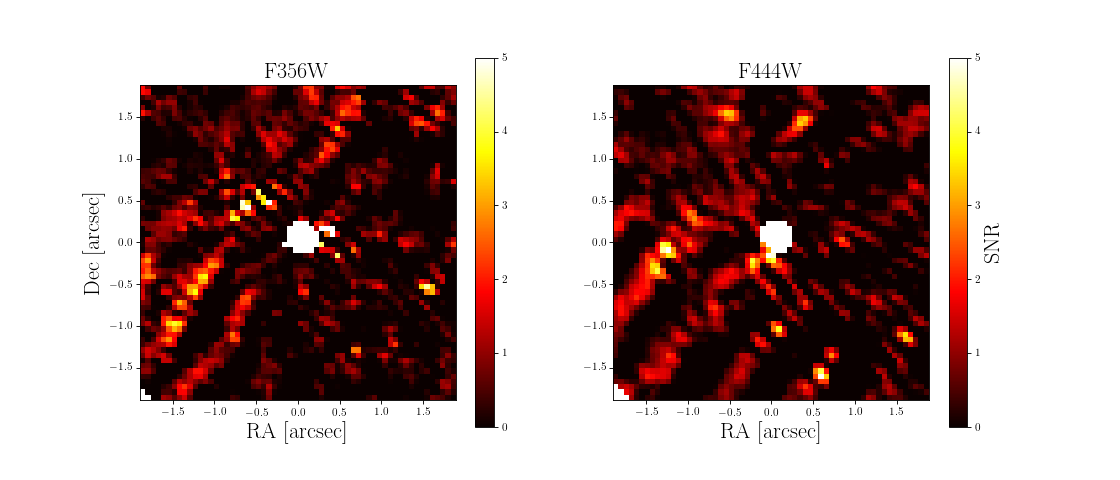}
  \label{fig:bottom_pca}}
\caption{(Top) PCA reduction of the inner, speckle-dominated, region (inside $\sim$1.5\arcsec).  To emphasize the better results in the F444W filter, the images are displayed with the same absolute scale.  The residual noise level is worse in the F356W image due to the brighter stellar flux and shorter integration time. (Bottom) SNR map from a forward model match filter (FMMF) analysis of the inner region of the Fomalhaut data. Pixels that yield an SNR of $\sim$5 are associated with residual bad pixels identifiable in the PSF-subtracted images and/or with residual instrument-related striping (diagonal structures going from the upper right to lower left). There are no statistically significant peaks that could be ascribed to a planetary companion.
\label{fig:innerRegionPCA}}

% \end{figure*}

% \begin{figure*}[t!]%
%      \centering
%      \begin{subfig}[b]{0.9\textwidth}
%          \centering
%          \includegraphics[width=0.9\textwidth]{fomalhaut_pca_allFilters.png}
%          \label{fig:top_pca}
%      \end{subfig}
%      \hfill
%      \begin{subfigure}[b]{0.9\textwidth}
%          \centering
%          \includegraphics[width=0.9\textwidth]{fomalhaut_fmmf_allFilters.png}
%          \label{fig:bottom_pca}
%      \end{subfigure}
%      \hfill
% \caption{(Top) PCA reduction of the inner, speckle-dominated, region (inside $\sim$1.5\arcsec).  To emphasize the better results in the F444W filter, the images are displayed with the same absolute scale.  The residual noise level is worse in the F356W image due to the brighter stellar flux and shorter integration time. (Bottom) SNR map from a forward model match filter (FMMF) analysis of the inner region of the Fomalhaut data. Pixels that yield an SNR of $\sim$5 are associated with residual bad pixels identifiable in the PSF-subtracted images and/or with residual instrument-related striping (diagonal structures going from the upper right to lower left). There are no statistically significant peaks that could be ascribed to a planetary companion.
% \label{fig:innerRegionPCA}}

\end{figure*}

\begin{figure*}[t!]
\centering
\includegraphics[clip,width=0.8\textwidth]{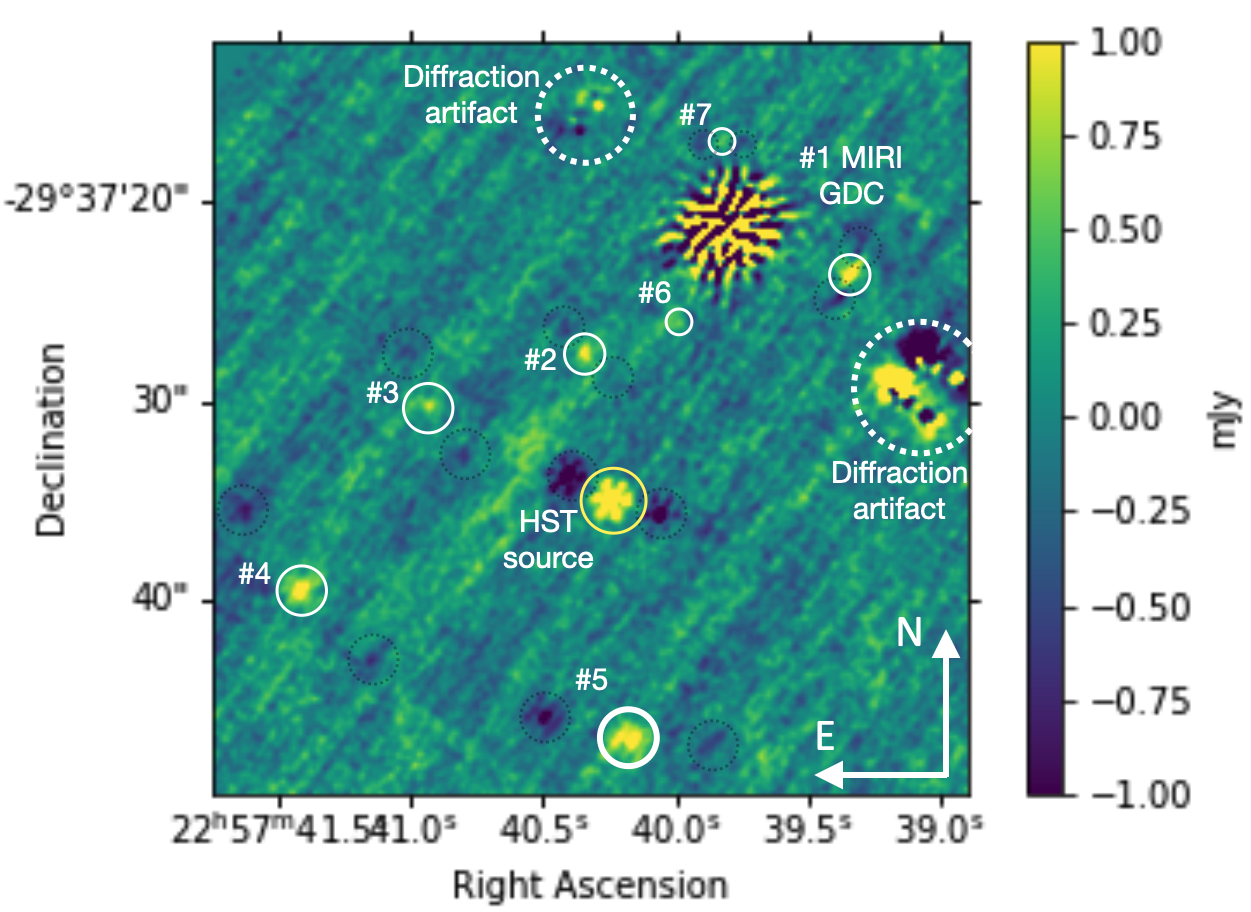}
% \includegraphics[clip,width=1.\textwidth]{F335+F444data_image.pdf}
%{figures%/data_NIRCam_image_F444W+F356W_FULL_zoom_final_paper_gf_classical_psfsub_with_sources.png} 
% \vskip -0.8in
\caption{
A zoomed-in F356W+F444W composite image highlighting the   NIRCam sources (\textit{S1-S7} and one HST source (see Table~\ref{allsources} for details). The negative-positive-negative pattern reflects the effects of the roll subtraction. The image was Gaussian-filtered with a $\sigma$=2 pixel kernel.  %Also shown are the identification numbers of the sources detected in Keck deep imaging \citep{Kennedy2023}. 
\label{fig:reduced_images_zoom_with_sources}}
\end{figure*}

% \begin{figure*}[t!]
% \centering
% % \includegraphics[clip,width=0.85\textwidth]{Figure_with_Keck_Sources.png}
% \includegraphics[clip,width=1\textwidth]{Figure_with_Keck_Sources_3.png}
% % \vskip -0.4in
% \caption{A combined F356+F444W full frame classical ADI-reduced image showing some of the sources detected with NIRCam (labelled \textit{S1-S7}) within and adjacent to the debris disk system. As discussed later in the text ($\S$\ref{results}) the figure also  shows sources identified in Keck H-band imaging \citep{Kennedy2023} obtained in 2005-2011 when \apsa\ was $\sim 5$\arcsec\ from the position at which the NIRCam data were obtained (Epoch=2022.808).
% \label{fig:igure_with_Keck_Sources}}
% \end{figure*}

\begin{figure*}[t!]
\centering
\includegraphics[clip,width=1\textwidth]{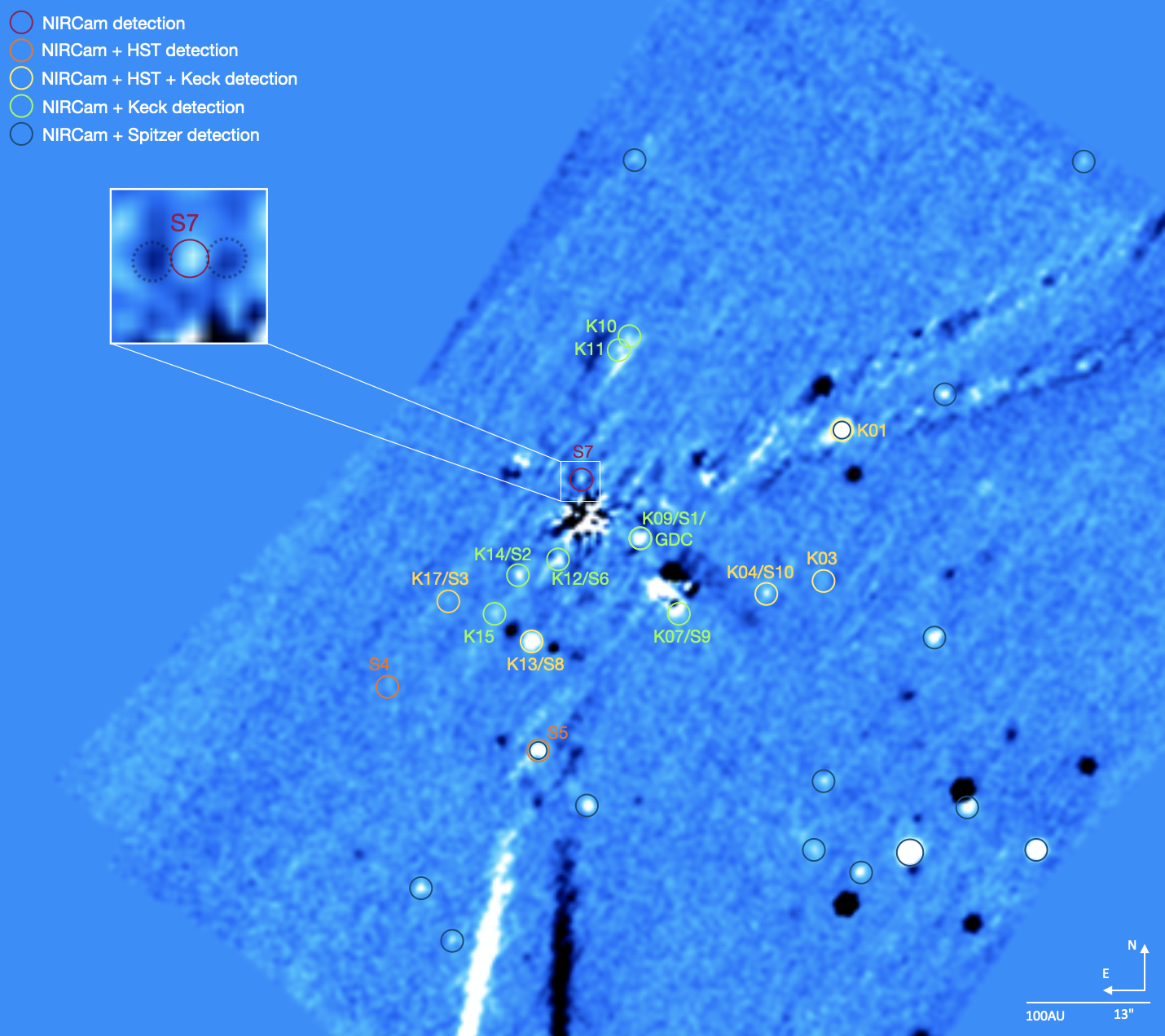}
% \vskip -0.4in
\caption{A combined F356+F444W full frame classical ADI-reduced image showing some of the sources detected with NIRCam (labelled \textit{S1-S7}) within and adjacent to the debris disk system. As discussed later in the text ($\S$\ref{results}) the figure also  shows sources identified in Keck H-band imaging \citep{Kennedy2023} obtained in 2005-2011 when \apsa\ was $\sim 5$\arcsec\ from the position at which the NIRCam data were obtained (Epoch=2022.808). Also shown are background sources detected with Spitzer 4.5 \mum\ imaging from 2006.9 \citep{Marengo2009}. 
\label{fig:igure_with_Keck_Sources}}
\end{figure*}

%\begin{figure*}[t!]
%\centering
%\includegraphics[clip,width=0.8\textwidth]{data_NIRCam_image_F444W+F356W_FULL_zoom_final_paper_gf_classical_psfsub_miri_contours.png}
% \vskip -0.4in
%\caption{Zoomed-image miri contour.
%\label{fig:reduced_images_zoom_with_miri_contours}}
%\end{figure*}

After PSF subtraction, a set of additional bad pixels became apparent in the inner, speckle dominated, region. The brightness of the speckles help these bad pixels elude correction in the first steps of correction; we, thus, need to correct for these after PSF subtraction (\S\ref{PSFsubtraction}). 
For the inner region ($\sim$3.7\arcsec$\times$3.7\arcsec) of the PSF-subtracted non-coadded images,
we apply a temporal and spatial bad pixel correction identical to the procedure described in the previous paragraph. 

To remove horizontal striping resulting from 1/f noise, we subtracted off the median value from each row of data (the direction of fast detector readouts). Figure~\ref{fig:raw_data} shows single F356W frames before and after horizontal stripe removal. 
Figure~\ref{fig:coadd_data} shows the final coadded images in both F356W and F444W, prior to the next post-processing steps to remove residual starlight in the images. 

\subsection{Point Spread Function (PSF) Subtraction}\label{PSFsubtraction}

Following the basic data reduction for individual images, we combined the overall set of two roll angles on the primary target and a set of dithered observations of the reference star, using either %\textbf{Reference star Differential Imaging (RDI) or Angular Difference Imaging (ADI)}.
Reference star Differential Imaging (RDI) or Angular Difference Imaging (ADI). We used two approaches for this post-processing --
classical PSF subtraction and a principal component analysis (PCA) taking advantage of the full diversity of the dithered reference PSFs.
For the classical RDI, %\textbf{RDI}, 
we first created a reference PSF from the nearby star HR 8709, shifting and coadding its five dithered observations together to maximize SNR.
We scaled and shifted the reference star PSF to align with the target at Roll~1 and Roll~2 independently before performing the PSF subtraction. 
For the classical ADI, we subtracted the two rolls from one another after applying the corresponding shift and data centering.  
In both RDI %\textbf{RDI} 
and ADI approaches, the last step after PSF subtraction was to orient both subtracted rolls to the North before coadding them resulting in a negative-positive-negative pattern for sources that are present in both telescope angles. 

While the classical PSF subtraction performs well at larger distances from the target star where the noise is limited by the instrument sensitivity, at close separations residual starlight speckles are the dominant limitation to the detection of point sources. PCA analysis is preferred for cleaning the inner speckle field within $\sim$1.5\arcsec. In this speckle dominated region %\textbf{we performed} 
we performed a PCA-based algorithm \citep{Amara2012} via Karhunen Lo\'eve Image Projection \citep[KLIP;][]{Soummer2012} using both the reference frames and the target frames from the opposite roll in the PSF library. %\textbf{using both the reference frames and the target frames from the opposite roll in the PSF library}. 
We used the open source Python package \texttt{pyKLIP} \citep{Wang2015}, which provides routines for cleaning the images, calculating detection limits, and quantifying the uncertainty in their flux of any detected sources.  

The KLIP reduction was done using all available images (i.e.\ with 6 KL-modes), with the full array mode data cropped and centered to match the subarray mode data. Only the full array dataset was used to produce the PCA full-frame reductions. The reference star data was used for  PSF subtraction. Use of the 5-POINT-SMALL-GRID dither pattern mitigated any  misalignment between  the star and coronagraph focal plane mask.  %The KLIP reduction parameters were as follows: The inner and outer angle limits are 2 and 41 pixels, this region is divided in 3 annuli, the number of basis is 6, i.e.\ the maximum number of basis given the two rolls plus the 5 reference star images.

Figures~\ref{fig:reduced_images_cADI},%~\ref{fig:reduced_images},
~\ref{fig:innerRegionPCA},~\ref{fig:reduced_images_zoom_with_sources}, and~\ref{fig:igure_with_Keck_Sources} show the results of the classical PSF and PCA subtractions.
The presence of the expected negative-positive-negative image pattern is a good indication that a candidate object is real, even if its overall SNR is too low for reliable extraction of its position and brightness. Sources \textit{S1-S10} discussed in Section~\ref{sources_detection} all show this pattern. 

\subsection{Contrast Calibration}

The contrast limits reported in this work were obtained by normalizing the flux to a synthetic peak flux. Predicted fluxes in the JWST wavebands were calculated based on BOSZ stellar models \citep{bohlin17} fit to optical and near-IR photometry (Table~\ref{tab:star}). Convolving the stellar model with JWST bandpasses gives an estimated flux of  115.6 Jy (0.93 mag) in the F356W filter and 80.9 Jy (0.89 mag) in F444W. To estimate the peak flux of the instrument's off-axis coronagraphic PSF we simulated this PSF using \texttt{WebbPSF} \citep{perrin14}. 
Measured fluxes in the NIRCam images were divided by these stellar fluxes to obtain contrast ratios.
%\textbf{The brightness of Fomalhaut produces nearly saturated near-IR photometry from 2MASS and WISE, resulting in relatively high uncertainty in the stellar flux estimates: $\sim$5\%; 
% \textbf{While the brightness of Fomalhaut often saturates CCD detectors at near-IR wavelengths (e.g.\ in the 2MASS and WISE all-sky surveys), \cite{carter1990} was able to use the 75-cm South African Astronomical Observatory to obtain JHK photometry with 2\% accuracy (see Table \ref{tab:star}), resulting in $\sim$2\% overall uncertainty in our stellar flux estimates at JWST wavelengths; 
% note that this uncertainty only contributes to the error budget for contrast ratios, not for the measured photometry that is used to calculate the contrast ratios.}
While the brightness of Fomalhaut often saturates CCD detectors at near-IR wavelengths (e.g.\ in the 2MASS and WISE all-sky surveys), \cite{carter1990} was able to use the 75-cm South African Astronomical Observatory to obtain JHK photometry with 2\% accuracy (see Table \ref{tab:star}), resulting in $\sim$2\% overall uncertainty in our stellar flux estimates at JWST wavelengths; 
note that this uncertainty only contributes to the error budget for contrast ratios, not for the measured photometry that is used to calculate the contrast ratios.

\begin{figure*}[b!]
\centering
\includegraphics[width=0.49\textwidth]
{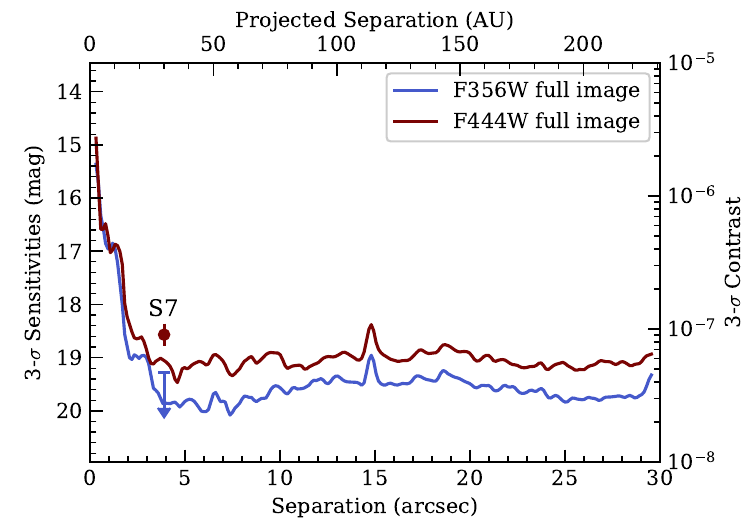}
\includegraphics[width=0.49\textwidth]{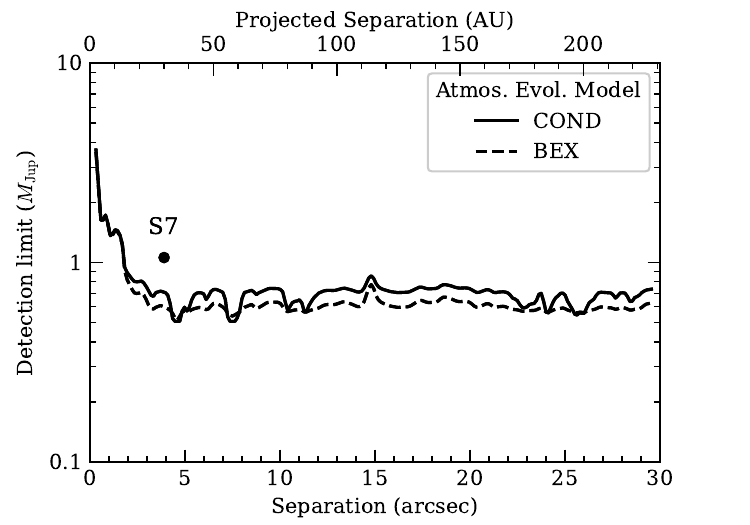}
\caption{
Coronagraphic observations of \apsa with NIRCam achieve a 3-$\sigma$ sensitivity below 10$^{-7}$ contrast at large separations from the star, limited primarily by the background and integration time.
%The observed contrast levels are shown for detected sources inside (\sone, \stwo, \textit{S6}, and \textit{S7}) and outside (\textit{S3, S4}, and \textit{S5}) the debris ring (as discussed below, $\S$\ref{results}).
At an age of 440 Myr \citep{Mamajek2012}, this contrast limit translates to a detection limit below $1$\mj\ outside of 2\arcsec, based on either AMES-Cond \citep{Baraffe2003} or BEX-HELIOS \citep{Linder2019} evolutionary models.
\label{fig:contrast}}
\end{figure*}

The 3-$\sigma$ contrast curves (Fig.~\ref{fig:contrast}a) were obtained using \texttt{pyKLIP}.
%with a 3-$\sigma$ threshold for detection.  
%\geoff{
%Since our aim is to identify any potential source in the field, we adopt a relatively relaxed threshold for detection.  
While a 3-$\sigma$ threshold allows for the possibility of spurious detections,
we find below that the overwhelming majority of sources above this threshold (all but one) are also detected by other telescopes, validating this threshold as appropriate for identifying potential sources for follow-up analysis.
%Cross-comparison against images from other telescopes verifies that the maj
%Further validation of these sources will be obtained below via cross-matching against images from other telescopes and between JWST filters.
%}
The noise was computed in an azimuthal annulus at each separation of the reduced image (before any smoothing), %\textbf{of the reduced image (before any smoothing)}, 
and we used a Gaussian cross correlation (kernel size = 2 pixels) %\textbf{(kernel size = 2 pixels)} 
to remove high frequency noise. The contrast was calculated using the normalization peak value for the target star. We corrected for algorithmic throughput losses by injecting and retrieving fake sources at different separations. The contrast was also corrected for small sample statistics \citep{Mawet2014} at the closest angular separations. 
At separations closer than 2\arcsec\ the contrast is limited 
against the residuals from the PSF subtraction methods ($\sim4\times 10^{-7}$ at 1\arcsec, $\sim 1 \times 10^{-7}$ at 2\arcsec), and further than 2\arcsec\ the performance is limited by the background level ($\sim 19$\,mag in F444W), consistent with the expectations of the instrument given the exposure time. 
Fig.~\ref{fig:contrast}b converts  the sensitivity curves into detection limits in Jupiter masses appropriate to Fomalhaut's age and distance ($< 1$ \mj\ beyond 2\arcsec).

Losses due to the coronagraphic mask were taken into account in both the contrast curves and in the reported fluxes for any detected sources
(the following section), although this is a negligible effect outside of $\sim$1\arcsec.
% \textbf{Transmission losses from the Lyot stop and coronagraph mask substrate were also included in both the contrast calibration and the point source photometry (\S\ref{sources_detection} below).
%These losses depend on the bandpass filter; 
% The mask substrate throughput is 0.95 and 0.93 for the F356W and F444W filters, while the transmission of the Lyot stop wedge is 0.98 in both.}
Transmission losses from the Lyot stop and coronagraph mask substrate were also included in both the contrast calibration and the point source photometry (\S\ref{sources_detection} below).
%These losses depend on the bandpass filter; 
The mask substrate throughput is 0.95 and 0.93 for the F356W and F444W filters, while the transmission of the Lyot stop wedge is 0.98 in both.

% S2 fluxes Jorge May 2023
%F356W: 3.089e-06 (+1.423e-06 -2.684e-06) Jy
%F444W: 1.246e-06 +1.143e-06 -2.253e-06 Jy 

% GDC fluxes at 2300 and 2550:61$\pm$12   & 89$\pm$13
%fluxes and mags update 3/17/2023
\begin{deluxetable*}{cc|cc|cccccc}
\tablecaption{Sources detected by JWST/NIRCam
\label{allsources}}
\tablehead{\colhead{Source} & \colhead{Other } & 
%\colhead{HST$^a$} & 
\colhead{$\Delta\alpha$}  &	\colhead{$\Delta\delta$}          &	
\multicolumn{3}{c}{F356W} &	\multicolumn{3}{c}{F444W}     \\
& Images$^a$ & (\arcsec)  & (\arcsec)
& Peak SNR$^b$ & ($\mu$Jy) & (mag) & Peak SNR & ($\mu$Jy) & (mag) 
%& ($\mu$Jy) & ($\mu$Jy) 
}
\startdata
%  GDC & MIRI & $-$6.05$\pm${0.3}	 & $-$2.42$\pm${0.3} & & & &  & & \\ 
%GDC - \textit{S1} &  & $-$0.05$\pm${0.3}	 & $+$0.14$\pm${0.3} & & & &  & & \\ \hline
%\hline 
\textit{S1}  & MIRI$^c$,K09 & $-$6.10$\pm$0.01 & $-$2.28$\pm$0.01    &	3.7 & 7.37$^{+0.58}_{-0.59}$ &	18.9$\pm$0.12    & 3.7 & 5.66$^{+0.56}_{-0.52}$ & 18.8$\pm$0.15  \\
% revised w Jorge May20 values
\textit{S2} & K14 & $+$7.22$\pm$0.03	 &  $-$5.94$\pm$0.04  & 3.3    & 3.33$^{+1.36}_{-1.53}$	 & 19.7$\pm$0.6 & 2.3 & 1.36$^{+1.22}_{-1.25}$ & 20.3$\pm$0.7  \\
\textit{S3}$^{d}$ & MIRI,HST/STIS,K17 & $+$14.89$\pm$0.04  & $-$8.49$\pm$0.06 &  3.4 & 5.47$\pm$1.14 & 19.2$\pm$ 0.2 &  3.8 & 3.71$\pm$0.99 & 19.2$\pm$ 0.3 \\
\textit{S4}$^{d}$ & MIRI,HST/WFC3 & $+$21.29$\pm$0.04 & $-$17.23$\pm$0.04 &  4.2 & 3.17$\pm$1.27 & 19.8$\pm$ 0.4 &  4.1 & 3.31$\pm$1.30 & 19.4$\pm$ 0.4 \\
\textit{S5}$^{d}$ & HST/WFC3/Spitzer & $+$4.69$\pm$0.01   & $-$25.31$\pm$0.01 & 3.6 & 9.81$\pm$1.02 & 18.6$\pm$ 0.1 &  5.2 & 14.04$\pm$0.76 & 17.8$\pm$ 0.1 \\
\textit{S6}$^{d}$ & K12 & $+$2.80$\pm$0.03   & $-$4.59$\pm$0.03 & 2.0 & 5.77$\pm$1.08 & 19.2$\pm$ 0.2 & - & 2.21$\pm$1.10 & 19.8$\pm$ 0.4 \\
\textit{S7}$^{d}$ & N/A & $+$0.45$\pm$0.03   & $+$3.91$\pm$0.03 & - & $<$4.59 & $>$19.3 & 3.0 & 6.53$\pm$1.24 & 18.6$\pm$ 0.2 \\
% \hline
\textit{S8} & HST/STIS,K13 & $+$5.77$\pm${0.003} &	$-$13.29$\pm${0.002} & 25.8 & 60.56$^{+2.16}_{-2.17}$  & 16.6$\pm$0.07 & 26.7 & 47.01$^{+ 1.21}_{-1.25}$ & 16.5$\pm$0.05 \\
\textit{S9} & HST/STIS,K07 & $-$20.09$\pm${0.013} &	$-$8.29$\pm${0.011} & 6.0 & 7.48$^{+0.82}_{-0.86}$  & 18.9$\pm$0.2 & 8.1 & 6.42$^{+ 0.69}_{-0.67}$ & 18.7$\pm$0.2 \\
\textit{S10}  & HST/STIS,K04 & $-$28.21$\pm${0.001} & $+$9.74$\pm${0.001} & 31.7 & 82.36$^{+1.26}_{-1.25}$  & 16.3$\pm$0.03 & 26.8 & 70.13$^{+ 0.93}_{-0.92}$ & 16.1$\pm$0.02\\ \hline
\apsa b$^e$& HST & $-$9.377&  11.144 & - & $<$3.05 & $>$19.3 & - & $<$3.62 & $>$19.3 \\
\hline
\enddata
\tablecomments{
The position offsets are relative to current location of Fomalhaut  (Epoch=2022.808),
%Coordinates are relative to the central star, 
located at $\alpha$,$\delta$ = 
%22$^h$57$^m$39.046$^s$,$-$29$^o$37$^\prime$20.05\arcsec  (this is J2000)
%22$^h$57$^m$39.622$^s$,$-$29$^o$37$^\prime$23.81\arcsec  (this is with proper motion)
22$^h$57$^m$39.615$^s$,$-$29$^o$37$^\prime$23.87\arcsec,  %(this also has parallax)
including the effects of proper motion and parallax.
As denoted in Figure~\ref{NIRCam}, the offsets in earlier HST epochs are consistent with the star's proper motion.
\\
%MIRI GDC position from Andras incl distortion:
%22:57:39.18 -29:37:26.119 and Fomalhaut around 22:57:39.648 -29:37:23.898\\
Astrometric precision is based on the uncertainty in peak fitting; distortion contributes an additional systematic uncertainty of $\sim$10-20 mas.}
\tablenotetext{a}{Identification numbers for detected objects as seen in imaging by Keck \citep{Kennedy2023}, HST \citep{Currie2013}, and/or JWST/MIRI \citep{Gaspar2023}.}\tablenotetext{b}{Peak SNR is the signal-to-noise found by the initial peak finding routine (sensitive to single-pixel outliers), while the measured fluxes are based on either an MCMC-driven fit to the post-processed instrument PSF or, in the case of extended sources, from aperture photometry.}
\tablenotetext{c}{The GDC source in the MIRI image is located at $\Delta \alpha$,$\Delta\delta$ = $-$6.05$\pm${0.3}\arcsec,$-$2.42$\pm${0.3}\arcsec.  The offset between MIRI's GDC and NIRCam's \textit{S1} is $-$0.05$\pm${0.3}\arcsec, $+$0.14$\pm${0.3}\arcsec. }
%\tablenotetext{c}{Source with corresponding MIRI detection}
\tablenotetext{d}{Flux densities and uncertainties for this source are derived from aperture photometry, rather than PSF fitting.}
\tablenotetext{e}{Upper limit at the predicted position of \apsa b, if it is on a bound orbit. A similar limit applies at the position of \apsa b on an unbound orbit.}
\end{deluxetable*}

\subsection{Detection and Characterization of Point Sources near \apsa}\label{sources_detection}

% We adopted two methods to identify and characterize sources in the PSF-subtracted images.
% %We adopted two methods to identify and characterize sources in the PSF-subtracted images \textbf{whether they are located inside or outside of the speckle-dominated region.}  

% First, we examined the entire region interior and just exterior to the debris ring, which has a radius of %$\sim$20\arcsec = 
% 140 AU. Sources were initially detected using a Gaussian-smoothed image to search for $>3\sigma$ candidates. These candidates were examined visually to identify stellar diffraction and other image artifacts. 

% Second, we focused on the inner, speckle-dominated region, looking for planet candidates within ($\sim$1.5\arcsec) ($\sim$10 to 35 AU) of Fomalhaut. Taking advantage of the telescope stability to generate a well defined   PSF, we applied a forward model match filter \citep[FMMF;][]{Ruffio2017} method. FMMF corrects for the KLIP's over-subtraction effects with a forward model \citep{Pueyo2016}. This forward model is then used as a match filter to enhance the SNR of sources for which the spatial structure is well matched to the expected point source PSF. The PSF for the forward model is computed for each filter with \texttt{WebbPSF} using the OPD wavefront error map closest in time prior to the observations. \textit{No sources were reliably detected within this speckle-dominated region.}

We adopted two methods to identify sources in the PSF-subtracted images,
whether we considered the speckle-dominated or the outer region.
%\textbf{whether we considered the speckle-dominated or the outer region.} 
%We adopted two methods to identify and characterize sources in the PSF-subtracted images \textbf{whether they are located inside or outside of the speckle-dominated region.}  

First, we focused on the inner, speckle-dominated region, looking for planet candidates within ($\sim$1.5\arcsec) ($\sim$10 to 35 AU) of Fomalhaut. Taking advantage of the telescope stability to generate a well defined PSF, we applied a forward model match filter \citep[FMMF;][]{Ruffio2017} method. FMMF corrects for the KLIP's over-subtraction effects with a forward model \citep{Pueyo2016}. This forward model was then used as a match filter to enhance the SNR of sources for which the spatial structure is well matched to the expected point source PSF. The PSF for the forward model was computed for each filter with \texttt{WebbPSF} using the OPD wavefront error map closest in time prior to the observations. No sources were reliably detected within this speckle-dominated region. %\textbf{The FMMF method is most useful for point source detection in the speckle dominated region so we used a different method to identify sources in the outer region of the image.}
The FMMF method is most useful for point source detection in the speckle dominated region so we used a different method to identify sources in the outer region of the image.

Second, we visually examined the entire region interior and just exterior to the debris ring, which has a radius of %$\sim$20\arcsec = 
140 AU. Sources were initially detected using a Gaussian-smoothed image (kernel size = 2 pixels) %\textbf{(kernel size = 2 pixels)} 
to search for $>3\sigma$ candidates. These candidates were examined visually to identify stellar diffraction and other image artifacts.

%\st{Successful detection of sources with both techniques adds confidence in the robustness of the detection.}

% The photometry and astrometry of the detected sources were recovered either via an MCMC
% \citep[\texttt{emcee};][]{Foreman-Mackey2013}
% fit to the PSF-subtracted data using \texttt{pyKLIP} if they were point-like or via aperture photometry if they appeared slightly extended.
%Aperture photometry was performed with a 0.5\arcsec\ radius aperture (about 8 pixels), 

The photometry and astrometry of the detected sources were recovered via an MCMC
\citep[\texttt{emcee};][]{Foreman-Mackey2013}
fit to the PSF-subtracted data using \texttt{pyKLIP}. %\st{if they were point-like or via aperture photometry if they appeared slightly extended.}. 
% \textbf{Sources for which the MCMC fit was unsatisfactory were analyzed with aperture photometry.} 
Sources for which the MCMC fit was unsatisfactory were analyzed with aperture photometry.
For sources that manifest themselves differently from one method to another, the differences in the methodologies serve as a diagnostic for the source characteristics.  %\st{Bad pixels, for example, will be detected more strongly in a single-pixel-based S/N estimate (referred to subsequently as ``Peak SNR") than} \textbf{Point sources, for example, will be detected more strongly} 
Point sources, for example, will be detected more strongly with a method that matches the source to the predicted PSF like MCMC. %\textbf{like MCMC}. 
Extended structures, on the other hand, %\st{will be detected most strongly in images that have been convolved with a Gaussian filter (typically with $\sigma=2$ pixels) and} 
will have poor fits to point source models.

Aperture photometry was performed with a 0.25\arcsec\ radius aperture (about 4 pixels) at the same location as the MCMC fit, 
with background calculated within a 16- to 24-pixel annulus.
While this aperture size is large relative to the nominal telescope resolution, diffraction by the coronagraphic mask and Lyot stop results in a much broader PSF, with a large fraction of the flux dispersed to wider angles; only 20\%/25\% of the flux is enclosed within the aperture for the F356W/F444W filters.
%less than half of the flux from a point source is contained within the 0.5\arcsec\ aperture.
%the enclosed energy is still only 

We simulated the coronagraphic PSF with \texttt{WebbPSF\_ext} \citep{Leisenring2023}, deriving aperture corrections of 3.87 and 4.81 
(multiplicative factors on the flux) for F356W and F444W, respectively, for our chosen aperture.
%We apply aperture corrections of 2.59 and 2.92 for F356W and F444W, respectively.
We also measured the flux of each target within smaller and larger apertures (ranging from 0.1 to 1.0\arcsec), finding that most of our sources yield consistent fluxes independent of aperture size (after applying the aperture correction), as expected for point sources.  Some sources, however, have fluxes that rise significantly with aperture (e.g.\ S2 and S5), suggesting source sizes that are a fraction of an arcsecond.
We determined the photometric uncertainty by placing the aperture at locations throughout the background annulus and measuring the dispersion in fluxes (again multiplied by the aperture correction factors).
The standard aperture size (0.25\arcsec) was chosen to yield optimal S/N for point sources, but also served adequately for the sources that are moderately extended.
The measured fluxes and uncertainties are given in Table \ref{allsources},
with those determined via aperture photometry explicitly flagged.
%\textbf{The calibration uncertainty for NIRCam (not included in the Table  \ref{allsources} uncertainties) is currently estimated as $\sim$5\% \footnote{https://jwst-docs.stsci.edu/jwst-data-calibration-considerations}, much smaller than the photometric uncertainties for our low S/N detections.}
The calibration uncertainty for NIRCam (not included in the Table  \ref{allsources} uncertainties) is currently estimated as $\sim$5\% \footnote{https://jwst-docs.stsci.edu/jwst-data-calibration-considerations}, much smaller than the photometric uncertainties for our low S/N detections.

Source positions were measured relative to Fomalhaut in the detector frame and converted to celestial coordinates using the star's position at the time of observation (including proper motion and parallax). As described in \citet{Greenbaum2023}, the position of the star in the detector was computed by performing cross-correlations of the data with synthetic PSFs, computed using \texttt{WebbPSF}. We use the \texttt{chi2\_shift} functions in the \texttt{image-registration} Python package\footnote{https://image-registration.readthedocs.io/}. This method yields uncertainties of $\sim$7~mas, consistent with \cite{Carter2022}.

We applied up to  $\sim$30 mas  distortion corrections to the WCS coordinate frame, based on a distortion map derived from on-sky observation of a dense stellar field (jwst\_nircam\_distortion\_0173). The correction was performed with the \texttt{jwst} Calibration Pipeline \citep{jwst2022}. We estimated the astrometric accuracy to range from $\sim$10 mas for sources close to the central star up to $\sim$30 mas for sources outside of the Fomalhaut ring.

Images at F356W and F444W were treated separately and results presented in Table~\ref{allsources}. By design the integration time in the shorter wavelength filter was smaller than at the longer wavelength, making the sensitivity at F356W worse than at F444W. The primary purpose of the F356W observation was rejection of objects with typical stellar or galaxy-like SEDs. 

We report in Table~\ref{allsources} the ``Peak SNR" associated with the brightest pixel of a given source. This is reported to quantify how much the source visually stands out over the noise. However, when extracting the photometry using the joint photometry and astrometry model fit, as explained above, the error bars may differ from the Peak SNR for lower signal detections. This discrepancy is due to the detections being excessively noisy due to bad pixels or poorly subtracted speckles. The figures in the Appendix show the model fit results of the data and model selected point-like sources, e.g. \sone, as well as the MCMC corner plots for the astrometric and photometric fits. The photometry and astrometry error bars provide  the most complete accounting of the properties of the sources, whereas the Peak SNR describes how well a source  can be seen over the smoothed noise in the image.

\section{Results \label{results}}

\begin{figure*}[t!]
\centering
\includegraphics[width=1.0\textwidth]{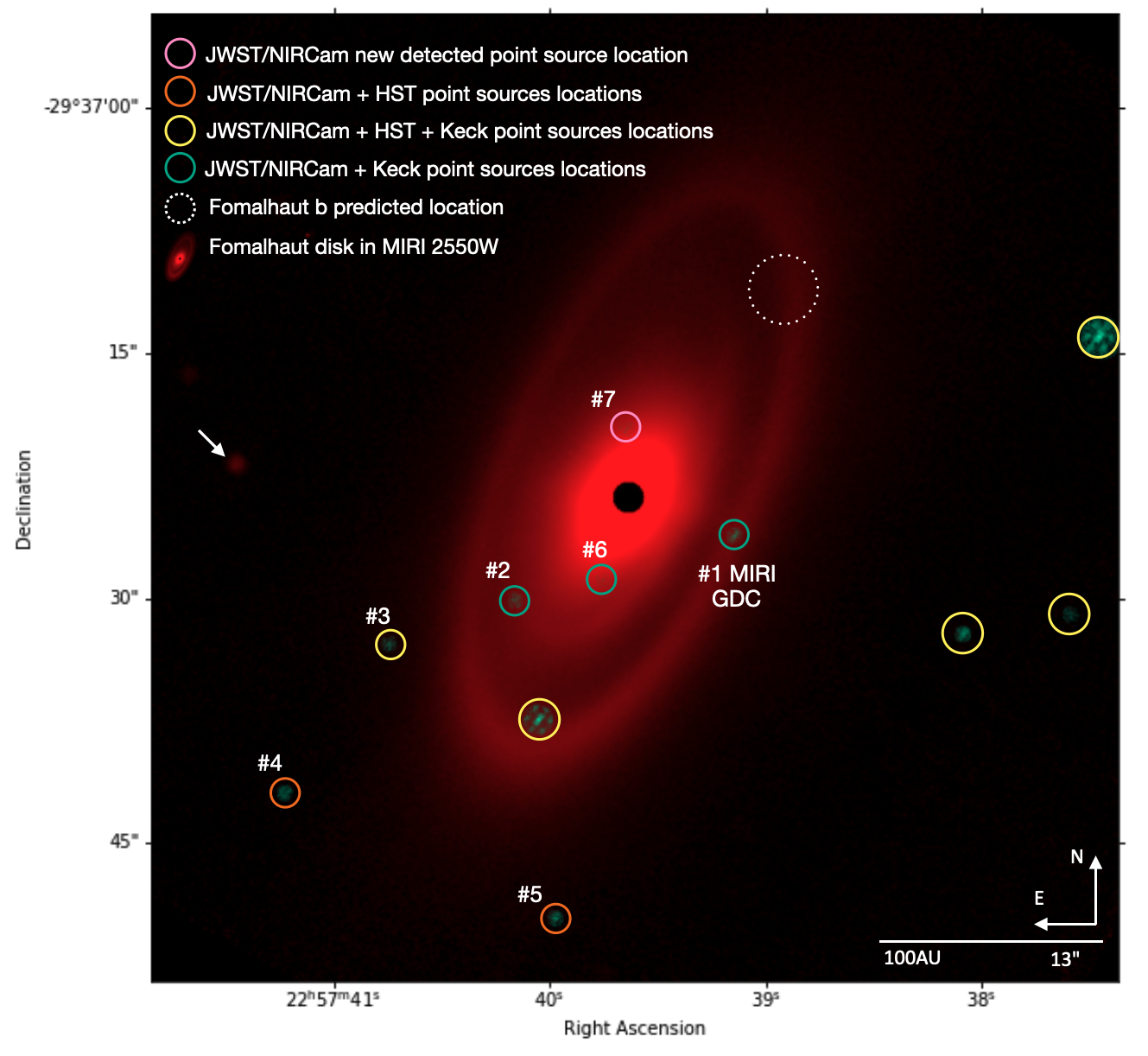}
% \vspace{-0.2in}
\caption{Point sources from the NIRCam/F356W+F444W image are shown with the  MIRI F2550W PSF-subtracted image superimposed in red. Extracted NIRCam sources without HST counterparts are shown as green circles; NIRCam sources with HST-detected background objects (yellow); and instrument artifacts associated with the edge of the coronagraphic mask field of view (white).  Source labels correspond to objects listed in Table~\ref{allsources}. The red dotted circles denote the expected position of \apsa b \citep{Gaspar2020}. No object is seen in this vicinity. We also note the presence of a bright MIRI point source at about 20\arcsec\ East of Fomalhaut that doesn't have a NIRCam counterpart. The location of this point source, marked
with a white arrow, coincides with the point source marked in Figure 6 of \cite{janson2015}. \label{NIRCam}}
\end{figure*}

\subsection{Sources in the Vicinity of \apsa }

Figure~\ref{fig:top_pca} shows the final images for the inner speckle field ($<$1.5\arcsec) around Fomalhaut. No sources are apparent, just a residual noise floor from imperfect speckle subtraction. Figure~\ref{fig:bottom_pca} shows the corresponding SNR map from the FMMF analysis of the inner region; applying an SNR=5 threshold, the data yield no detections. 

%dvanced Camera for Surveys (ACS) coronagraph in 2004 and 2006

Figure~\ref{NIRCam} shows the sources extracted from the final NIRCam reduced image, %over the full field, 
along with a superposition of the MIRI data at F2300C.  Some objects are clearly detected in Keck (circled in green), HST (circled in orange) or both HST and Keck (circled in yellow) images from earlier epochs.\footnote{See Table 1 of \citet{Gaspar2020} for a comprehensive list of HST observations of Fomalhaut, which include the Advanced Camera for Surveys (ACS) coronagraph in 2004 and 2006, and STIS in 2010 and  2014.} One source, identified with a white arrow in Figure~\ref{NIRCam}, is visible in the MIRI data but has no counterpart in the NIRCam data. The location of this source coincides with the red point source identified in Spizter data from 2013 (see Figure 6 of \cite{janson2015}), indicating a background object. It is possible the source is variable and not detectable at the NIRCam sensitivity levels, particularly at the very edge of the field of view. Alternatively, if the Spitzer object is not associated with the MIRI source and is indeed part of the Fomalhaut system, then the proper motion of Fomalhaut from 2013 to present time would have resulted with the object falling very close to the edge of the field of view of the current NIRCam observations. 

% Image artifacts are marked with dashed white circles. 
%NIRCam sources are circled in green.

Table~\ref{allsources} summarizes the properties of the objects detected within 30\arcsec\ of the star, corresponding to a projected separation of 230 AU. Ten point sources are detected at either or both F356W and F444W with greater than 3-$\sigma$ significance.
We have compared these sources against those detected in observations by Keck \citep{Kennedy2023}, HST \citep{Currie2013}, and JWST/MIRI \citep{Gaspar2023};
all but one source -- S7 -- have been previously imaged.
We find that none of these previously identified sources are co-moving with Fomalhaut and we conclude that they are background objects.

A brief description of each source follows:
\begin{itemize}

\item \textit{S1}: The source sits within the combined NIRCam-MIRI positional uncertainties at the clump of emission seen in the MIRI coronagraphic image at 23.0 and 25.5 $\mu$m and denoted as the Great Dust Cloud (or GDC) by \citet{Gaspar2023}. Detected at both F356W and F444W, it is separated from the GDC by $(\alpha,\delta)=(-50\pm300,+140\pm300)$ mas where the dominant source of uncertainty comes from the lower resolution MIRI image. The F356W/F444W color, [F356W]--[F444W]=0.0$\pm$0.2 Vega mag), is relatively blue (corresponding to a color temperature of $\sim$1700$\pm$400 K).  There is a Keck object, \textit{K9}, as well as an ALMA source seen at this position confirming this as a background object \citep{Kennedy2023}.  Astrometric coincidence  within $\sim 100$ mas  with Keck H-band objects from  earlier epochs (2005-2011) is used to identify \sone\ and  all but one of the other NIRCam detections discussed below as background objects  ($\S$\ref{discuss}). The F356W data for \sone\ show a hint of being extended, consistent with the the galaxy interpretation (Figures~\ref{fig:reduced_images_zoom_with_sources},~\ref{fig:mcmc1}). As discussed further below ($\S$\ref{discuss}), comparing the combined NIRCam/MIRI spectral distribution with Spitzer SWIRE templates \citep{Poletta2007} suggests that the object is an  ultra-luminous IR galaxy like Arp 220 or M82 at a redshift $z$=0.8-1.0 

\item \textit{S2}: This source is clearly detected at F356W with a peak SNR of 3.3 and marginally in F444W (Figure~\ref{fig:mcmc2}). Although \stwo\ is positioned tantalizingly close to the outer edge of the innermost disk, close to the gap between the disk and the intermediate belt seen in the MIRI images \citep{Gaspar2023}, the presence of a Keck source at this position, \textit{K14}, makes association with \apsa\ improbable.
\begin{figure*}
\centering
\includegraphics[width=0.99\textwidth]{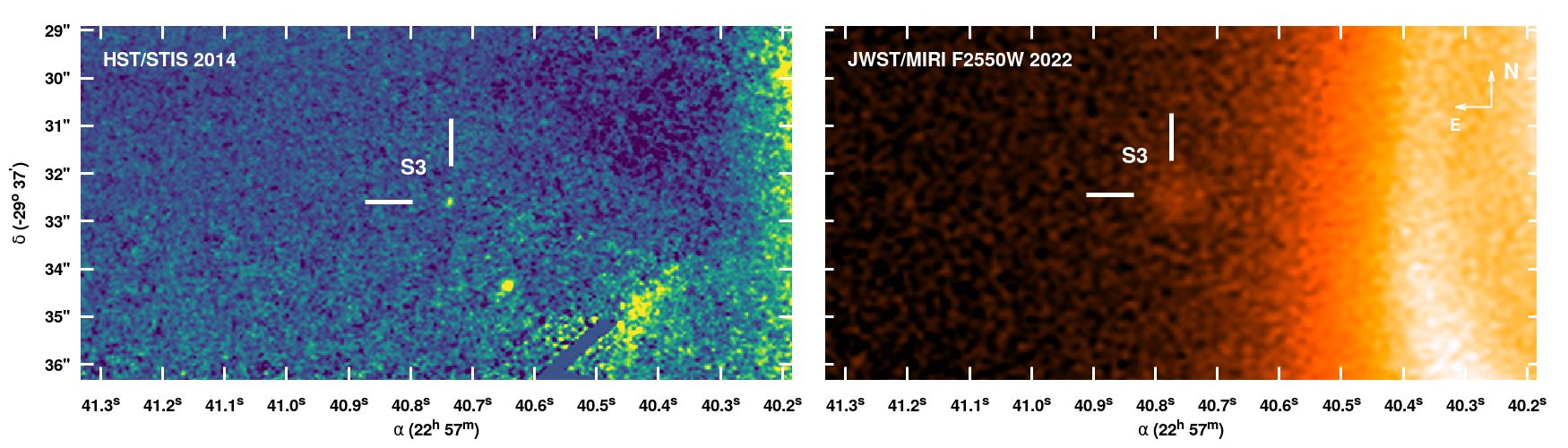}
\caption{A HST/STIS image from 2014 (left) and MIRI F2550W (right) showing the detection of \textit{S3} in both the HST and  MIRI data. The source
is detected at $\alpha=$22$^{\rm h}$57$^{\rm m}$40$\fs$73, $\delta=-29^{\circ}37^{\prime}32\farcs$63 in the HST/STIS image and offset by $0\farcs39$ in the MIRI image at $\alpha=$22$^{\rm h}$57$^{\rm m}$40$\fs$76, $\delta=-29^{\circ}37^{\prime}32\farcs$53.\label{fig:STIS}}
\end{figure*}
    
% Source #3:I think this is absolutely detected in the 2014 STIS data (not in the others). Small positional offset uncertainty. STIS flux (~ 0.6 micron): 0.312 μJy (25.16 ABmag, 24.92 VegaMag)F2550W flux: 23.2 +/- 3.91 mJy

\item  \textit{S3 \& S4}: These objects have  high peak SNRs in  both F356W  and F444W while the point source MCMC analysis provides a poor fit to the data. Aperture photometry of a slightly extended source ($\sim$0.25\arcsec-0.5\arcsec) provides  robust photometric detections in  both bands for both objects. \textit{S3} is detected in the Keck deep imaging, the 2014 HST/STIS image and in the MIRI data at F2550W \citep[Figure~\ref{fig:STIS}]{Gaspar2023}.  Although \textit{S4} is not seen in  Keck imaging, it is detected by HST/WFC3 and its  relatively blue color ([F356W]-[F444W]$\sim$ 0 (Vega mag) is consistent with its being a distant galaxy. \textit{S4}  is not seen in the F2550W imaging and falls outside the field of view of the F2300W data.

\item  \textit{S5}: This object sits well outside the \apsa outer ring. It is robustly detected at both both F356W  and F444W. It is not seen in the Keck imaging and falls outside of the field of view of available HST data. Although the source  is well fit with the PSF MCMC analysis (Figure~\ref{fig:mcmc5}), the aperture photometry finds increasing flux for larger apertures, suggesting an extended object.  Although the [F356W]-[F444W]=0.8$\pm$0.2 mag  (Vega) color is more typical of a distant brown dwarf than a field galaxy ($\S$\ref{discuss}), the hint of extended emission suggests the latter interpretation. 

\item  \textit{S6}:   This object has  marginal detections at F356W and F444W but is spatially coincident with an object seen in deep Keck imaging (\textit{K12}). With  [F356W]-[F444W]=0.6$\pm$0.4 (Vega mag) it  likely to be a background galaxy.

\item  \textit{S7}: This object is a $\sim$5$\sigma$ detection at F444W and not detected at F356W, giving it a  [F356W]-[F444W] color $\geq$0.7 (Vega mag). It is without  a Keck counterpart.  If it were an exoplanet, it would have a mass $\sim$1\mj\ as suggested by Figure~\ref{colorMagFig} and various models  \citep{Baraffe2003, Linder2019}. What is most intriguing about this object, the only NIRCam object that cannot  be immediately associated with a background source, is its  proximity to the  inner dust disk newly identified in the MIRI imaging \citep[Figure~\ref{NIRCam}]{Gaspar2023}. This disk extends from $>$1.2\arcsec\ outward to 10-12\arcsec, compared with the location of \textit{S7} at 4\arcsec\ ($\sim$ 30 AU) separation from Fomalhaut. If associated with Fomalhaut and of the indicated mass, it should have substantial dynamical interactions with the inner debris disk, which are not evident in the 25.5 $\mu$m image. It will be important to address its possible effects on the structure of the inner disk if its planetary nature is confirmed. 

\item  \textit{S8, S9 \& S10}: These three objects are easily detected in both earlier HST/Keck imaging and the NIRCam data, making them obvious background objects. In particular, \textit{S10} looks extended in HST images. 
\end{itemize}

Figure \ref{colorMagFig} compares the magnitudes and colors of our detected point  sources.  
%\textit{S1-S6} 
Most sources have 0 $<$ [F356W]--[F444W] $< 1$ Vega mag (or 0.7$\pm$0.5 AB mag), 
consistent with typical galaxy colors at this sensitivity level \citep[Figures 30 \& 31 in][]{Ashby2013, Bisigello2023}. 
The newly identified source \textit{S7}, however, only has an upper limit on its [F356W]--[F444W] color.
For comparison with \textit{S7} we show planet evolutionary curves from  AMES-Cond \citep{Baraffe2003} and BEX \citep{Linder2019}. For the BEX cooling curves, we consider two different radiative transfer models for the planetary atmosphere, HELIOS and the version of petitCODE with clouds \citep{Linder2019}. With only an upper limit to the brightness of this object at F356W, we cannot make a definitive statement about the nature of \textit{S7}, but based on its brightness at F444W alone, its mass is at or below 1\mj.
% \textbf{Note that there is no sign of the planet predicted by \cite{Janson2020} as our detection limit (19 mag or $\sim$ 330 M$_\mathTerra$ in F444W) is higher than the one they had predicted for those observations (24 mag or $\sim$ 66 M$_\mathTerra$ in F444W).}
Note that there is no sign of the planet predicted by \cite{Janson2020} as our detection limit (19 mag or $\sim$ 330 M$_\mathTerra$ in F444W) is higher than the one they had predicted for those observations (24 mag or $\sim$ 66 M$_\mathTerra$ in F444W).

\begin{figure*}[t!]
\centering
\includegraphics[width=0.6\textwidth]{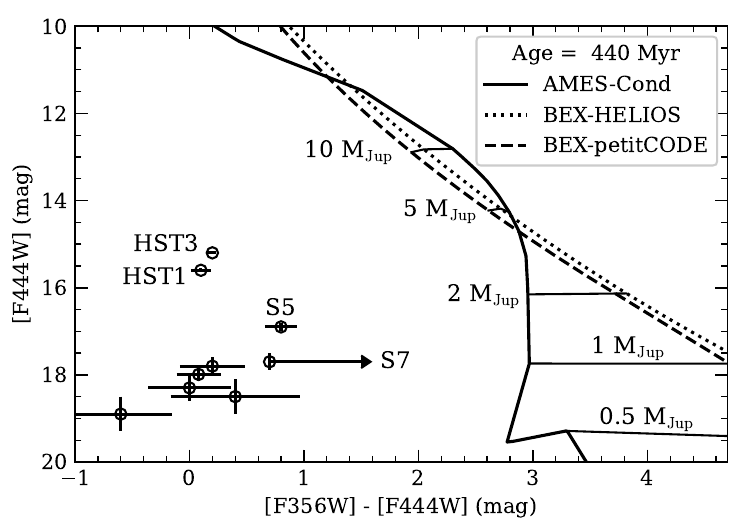}
\caption{ 
The F444W brightness and [F356W]$-$[F444W] color of our detected sources 
are compared against 440 Myr isochrones as a function of planet mass. Three evolutionary-radiative models are considered --
AMES-Cond (solid line), BEX-HELIOS (dotted), and BEX-petitCODE with clouds (dashed).
Masses from 0.5 to 10 \mj\ are marked on each isochrone. In the lower left hand corner the source IDs are omitted for clarity, but from left to right are: \textit{S6}, \textit{S3}, \textit{S1}, \textit{S9}, and \textit{S4}.
\label{colorMagFig}}
\end{figure*}

\subsection{What About Fomalhaut b?}

The expected position of \apsa b at the time of the JWST observations depends on whether the object is on a bound or unbound orbit \citep{Gaspar2020}. The expected offsets with respect to \apsa are: ($\Delta\alpha,\Delta\delta)=$ ($-$9.377\arcsec, 11.144\arcsec) and    ($-$9.809\arcsec, 11.665\arcsec), for the bound and unbound orbits,  respectively. The failure to  detect any F356W or F444W emission at the  predicted position(s) of \apsa b rules out the presence of any object more massive than $\sim$1\mj\ and is consistent with the hypothesis that the object is a dispersing remnant of a collision between two planetesimals. The scattered light seen at shorter wavelengths by HST would be much fainter at NIRCam wavelengths due to lower stellar flux ($\propto \lambda^{-2}$) and the expected lower scattering cross sections.
\citet{Janson2020} suggest that the disruption of a planetesimal in the tidal field of a  $\sim$0.2 \mj\ planet located at  semi-major axis of 117 AU might be the cause of the observed collisional remnant. Such an object is below the sensitivity of the current observation.

\subsection{Upper Limits on Scattered Light from the Debris Disk}

%{\color{red}{Mike's comment: Need some more references here, particularly w.r.t  the albedo as a function of particle size and composition (and wavelength).}} 

We do not detect scattered light from the outer debris ring. Starting with the RDI-processed images (not ADI images, where the extended disk would self-subtract during the roll subtraction),
we summed the flux within an ellipsoid annulus corresponding to the known ring location, but do not find any emission above the background.
Based on the azimuthal rms-deviation between 30\degree\ bins,
we find an upper limit on the disk emission of $\sim$17 $\Delta$mag/arcsec$^2$ in both the F356W and F444W filters. This contrast is more than an order of magnitude less than the 
optical contrast detected by HST/ACS in combined F606W and
F814W data %\textbf{HST/ACS in combined F606W and F814W data} 
\citep[$\sim$20 $\Delta$mag/arcsec$^2$;][]{Kalas2005}. 
While JWST/NIRCam has lower effective throughput than HST/ACS (partially negating the advantage in primary mirror size), the poorer contrast limit relative to HST is primarily a result 
of fainter stellar flux and lower scattering cross sections at the longer wavelengths.  Typical interstellar grains have  a factor of $\sim$10 smaller albedo at NIRCam wavelengths than at HST wavelengths \citet{Draine2003} although large ($a$=3-5 \mum), ice-dominated  grains can have comparable  visible and near-IR albedos around 3-4\mum\  \citep{Tazaki2021,McCabe2011}. 

Comparing the JWST upper limit against the integrated thermal emission for the dust ring \citep[$L_{\rm disk}/L_{\star} = 5.4 \times 10^{-5}$;][]{Acke2012}, we place an upper limit on the dust albedo of $<$0.6 at JWST wavelengths (3.56, 4.44 \mum). While this rules out grains of pure ice, such dust has already been excluded
by the much lower HST albedo measurement of $\sim$0.05-0.10 at optical wavelengths.

\section{Discussion\label{discuss}}

%\subsection{Background Contamination\label{background}}

All of the objects in Table~\ref{allsources} with the  exception of \textit{S7} have counterparts in deep Keck or HST imaging from earlier epochs. The incidence of background objects (almost exclusively galaxies at these wavelengths and sensitivity levels)  can be assessed from a variety of references. \citet{Hutchings2002} suggest 10 objects  per sq.\ arcmin down to K$_s$=20.5 mag (Vega). \citet[Figures 32 \& 33]{Ashby2013}  find cumulative source densities of 15 sources per sq.\ arcmin down to [IRAC2] = 19 (Vega mag) or 22.2 (AB mag) and 24 sources per sq.\ arcmin down to [IRAC1] = 20 (Vega mag) or 22.8 (AB mag). At longer wavelengths, \citet{Papovich2004} find a cumulative source density of $\sim$8 per sq.\ arcmin down to the 60 $\mu$Jy brightness of the MIRI GDC cloud at 23 \mum.  A rectangular region containing  the entire MIRI disk is approximately 40\arcsec$\times$20\arcsec=0.22 sq.\ arcmin, leading to an expected number of 3$\sim$5 F444W sources compared to the $\sim$7 seen here. The projected annular size of the outer ring itself is smaller, 0.06 sq.\ arcmin, with an expectation of $\sim$1 source  within the outer ring.

\begin{figure}[t!]
\centering
\includegraphics[width=0.85\textwidth]{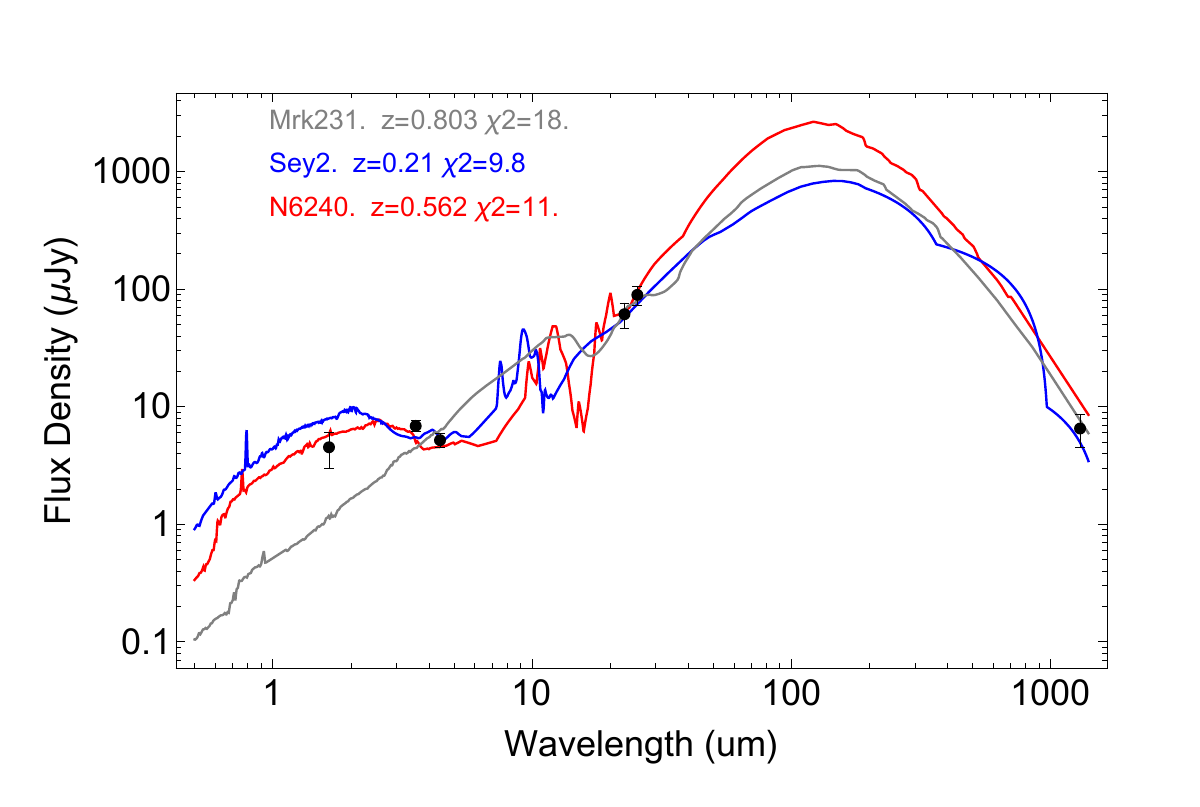}\\
\caption{ The  spectral energy distribution (SED) of \sone\ fitted to the model SEDs for three archetypal active galaxies,  Markarian 231 (grey), average Seyfert 2 (blue) or NGC6240 (red) at redshifts of $z=$(0.80,0.21 and 0.56, respectively) using templates from the Spitzer SWIRE catalog \citep{Poletta2007}\label{Arp220}}
\end{figure}

\citet{Poletta2007} give template SEDs covering visible to far-IR wavelengths   for 25 galaxy types, including ellipticals, spirals, AGN, and starburst systems\footnote{http://www.iasf-milano.inaf.it/~polletta/templates/swire\_templates.html}. These templates were derived  using the GRASIL code fitted to  data obtained between UV to far-IR wavelengths \citep{Silva1998}.  Varying only the redshift and an amplitude scaling factor, we fitted photometric data from  NIRCam,  MIRI, Keck and ALMA  for \sone. The Keck H-band brightness is 20.9$\pm$0.3 Vega mag or 4.5$\pm$1 $\mu$Jy  \  \citep{Kennedy2023}. We estimate the ALMA 1.3 mm flux density  as 5$\times$ the 1$\sigma$ noise level of 1.3 $\mu$Jy=6.5 $\mu$Jy\ \citep{MacGregor2017}.
%\textbf{we fitted photometric data from  NIRCam,  MIRI, Keck and ALMA  for \sone. The Keck H-band brightness is 20.9$\pm$0.3 Vega mag or 4.5$\pm$1 $\mu$Jy  \  \citep{Kennedy2023}. We estimate the ALMA 1.3 mm flux density  as 5$\times$ the 1$\sigma$ noise level of 1.3 $\mu$Jy=6.5 $\mu$Jy\ \citep{MacGregor2017}.}
The best fitting SEDs correspond  to find those of  active galaxies like Markarian 231, NGC6240 or the average Seyfert 2 (Figure~\ref{Arp220}). These luminous  starburst galaxies have the large amounts of  hot dust needed to  be consistent with the observations and which is lacking in the more normal spiral and other galaxies in the SWIRE library. Starburst galaxies of this type are quite common at  redshifts $\sim$1 \citep{Kartaltepe2012}.

% Similarly,  the STIS+NIRCam+MIRI data\footnote{{\color{red}The STIS flux  for \textit{S3} is (0.6 $\mu$m, 0.32$\pm$TBD $\mu$Jy) \citet{Gaspar2020} and the MIRI flux (25.55 $\mu$m, 23.2 $\pm$ 3.9 $\mu$Jy) \citet{Gaspar2023} } }  for \textit{S3}  yield an SED consistent  with that of a Sc galaxy at  $z=$1 (Figure~\ref{Arp220}b). 

Late-type T or Y brown dwarfs at distances of a few hundred parsecs are an alternative contaminant. Comparing the  [F356W] magnitudes and the [F356W]--[F444W] colors of \stwo\  with IRAC Ch1 and Ch2 colors of nearby brown dwarfs suggests that \textit{S5} could be a mid-T dwarf at 120 pc  \citep{Luhman2007}. Distant brown dwarf candidates have been found in the GLASS-JWST field \citep{Nonino2023} and in the JADES field \citep{Hainline2020} with multi-filter SEDs well fitted by T/Y brown dwarf models. %\textbf{Twenty-one} 
Twenty-one secure and possible brown dwarf candidates were identified in  the medium and deep JADES footprint   of 0.017 sq.\ deg, or 0.33 objects per sq.\ arcmin for temperatures ranging from  400 to 1400 K. This result is consistent with theoretical expectations of  0.2-0.4 M8-T8 objects per sq.\ arcmin at J$<$30 (AB mag) \citep{Ryan2016} at a galactic latitude similar to Fomalhaut's ($b=-65^o$). 

\textit{S7} has [F356W]-[F444W]$>$0.7 (Vega Mag) which is relatively rare among field galaxies  \citep[Figure 31]{Ashby2013} and lowers the probability that it is a background galaxy compared to the above numbers. At the same time the %\textbf{lower limit of the color} 
lower limit of the color is plausible for a distant T dwarf but  significantly  bluer than for typical models for Jovian mass objects (Figure~\ref{colorMagFig}). 
But \textit{S7}'s  color is only a limit and neither the background object nor the exoplanet explanation can be confirmed or rejected  in the absence of the  second astrometric epoch and improved photometry now approved for Cycle 2 (PID\#3925). 

%0.312 μJy (25.16 ABmag, 24.92 VegaMag)F2550W flux

%;   bottom) The STIS+NIRCam+MIRI  SED   of \textit{S3}  fitted to a model SED of an Sc spiral galaxy at a redshift of $z=$1.1 using a template the Spitzer SWIRE catalog.

\section{Conclusions}

NIRCam coronagraphy was used to examine the regions around \apsa in search of candidate planets that  might explain the structure of the debris disk with its multiple rings. The observations achieved contrast levels of $10^{-7}$ outside of 1\arcsec\ corresponding to planet masses $\leq$1 \mj. %No objects with appropriate colors and magnitudes were identified as obvious planet candidates. 
Nine of the ten reliable detections correspond to objects seen in Keck or HST imaging from earlier epochs, ruling them out as exoplanet candidates. The object  \textit{S7} is located within the inner dust ring seen in MIRI imaging and is detected only in  F444W. It has  has no obvious counterpart in earlier epochs and so might be a candidate planet subject to verification in future observations. 

The source \sone\ is directly associated with the MIRI GDC object has counterparts in NIRCam and earlier deep imaging. Its  SED is similar to ultra-luminous IR galaxies like Arp 220. The other objects similarly have NIRCam colors consistent with background galaxies, or possibly in the case of \textit{S5}, a distant brown dwarf. No NIRCam object is seen at the position of Fomalhaut $b$, consistent with its interpretation of a dissipating remnant of a collisional event.

 It is important to note that outside of the speckle dominated region, the sensitivity of these observations is limited by detector noise and the selected integration time. The approved JWST Cycle 2 program (PID \# 3925) will have almost 4 (F444W) and 8 (F356W)  times more integration time in full frame imaging  than this initial reconnaissance and will push the detection limit from $\sim$0.6\mj\ down to $\sim$0.3-0.4\mj\ at separations $\gtrsim$5\arcsec.  In addition to confirming (or rejecting) \textit{S7} as being associated with Fomalhaut, the Cycle 2 program might identify one or more of the planets expected to exist on the basis of the complex disk structure discovered in the MIRI results \citep{Gaspar2023}. %An upcoming deep HST imaging program (P. Kalas, PID \#17139) may also provide the valuable  information as to whether \sone,  \stwo, and possibly \textit{S5} are background objects or associated with \apsa. 

%\clearpage 

\section*{Acknowledgements}
%PUT BACK
% \begin{acknowledgements}
% \textbf{This work is based on observations made with the NASA/ESA/CSA James Webb Space Telescope. The data were obtained from the Mikulski Archive for Space Telescopes at the Space Telescope Science Institute, which is operated by the Association of Universities for Research in Astronomy, Inc., under NASA contract NAS 5-03127 for JWST. These observations are associated with program \#1193.
% NIRCam development and use at the University of Arizona is supported through NASA Contract NAS5-02105.
% A team at JPL's MicroDevices Laboratory designed and manufactured the coronagraph masks used in these observations. }
This work is based on observations made with the NASA/ESA/CSA James Webb Space Telescope. The data were obtained from the Mikulski Archive for Space Telescopes at the Space Telescope Science Institute, which is operated by the Association of Universities for Research in Astronomy, Inc., under NASA contract NAS 5-03127 for JWST. These observations are associated with program \#1193.
NIRCam development and use at the University of Arizona is supported through NASA Contract NAS5-02105.
A team at JPL's MicroDevices Laboratory designed and manufactured the coronagraph masks used in these observations.
Part of this work was carried out at the Jet Propulsion Laboratory, California Institute of Technology, under a contract with the National Aeronautics and Space Administration (80NM0018D0004). 
We are grateful for support from NASA through the JWST NIRCam project though contract number NAS5-02105 (M. Rieke, University of Arizona, PI).
The High Performance Computing resources used in this investigation were provided by funding from the JPL Information and Technology Solutions Directorate. 
The work of A.G.\, G.R.\, S.W.\, and K.S.\ was partially supported by NASA grants NNX13AD82G and 1255094. 
D.J.\ is supported by NRC Canada and by an NSERC Discovery Grant. 
We thank Dr. Paul Kalas for discussions about HST observations of Fomalhaut. 
William Balmer, Jens Kammerer and Julien Girard provided valuable hints about early stages of pipeline processing. 
% \end{acknowledgements}

\copyright 2023. All rights reserved.

\facilities{JWST}

\software{
\texttt{astropy} \citep{astropy2013,astropy2018,astropy2022},
\texttt{jwst} \citep{jwst2022},
\texttt{NIRCoS} \citep{Kammerer2022},
\texttt{pyNRC} \citep{Leisenring2023},
\texttt{pyKLIP} \citep{Wang2015},
\texttt{species} \citep{Stolker2020},
\texttt{WebbPSF} \citep{perrin14},
\texttt{WebbPSF\_ext} \citep{Leisenring2023}
}

\appendix

\section{\texttt{pyKLIP} Results of Forward Model Photometry and Astrometry} \label{sec:fmstamps}
\begin{figure*}[t!]
\centering
\includegraphics[width=1.15\textwidth,trim={70 0 0 0},clip]{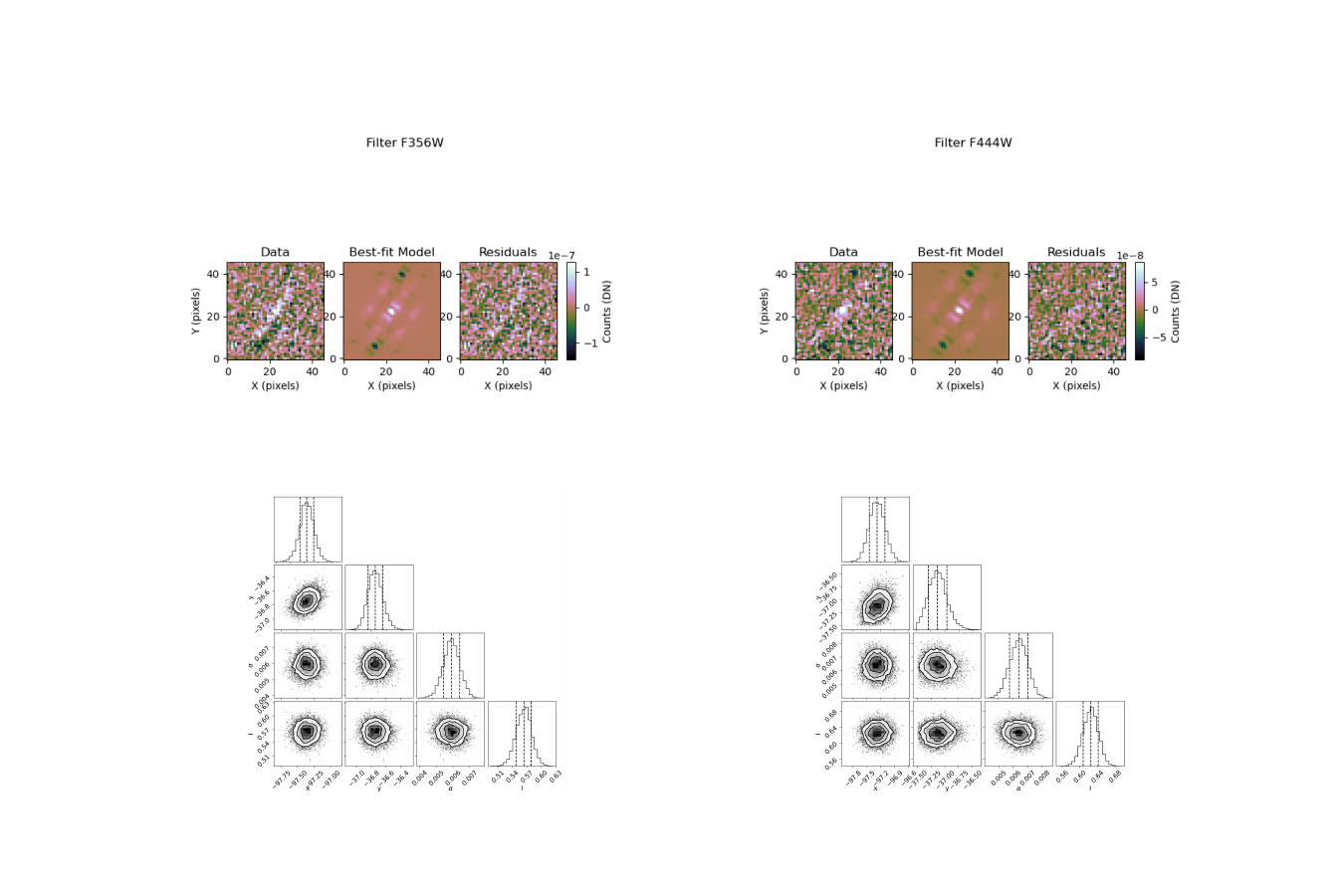}
\vskip -0.4in
\caption{ 
Top: Three panel images showing data, model and residuals for NIRCam \textit{S1}  at F356W (left) and F444W (right). 
Bottom:  The MCMC post-processing analysis gives (x,y) position in pixels, the relative flux, and a correlation length, a Gaussian hyperparameter.  This fit shows robust detections at both wavelengths. The same analysis is applied to all sources discussed in Table~\ref{allsources}. }
\label{fig:mcmc1}
\end{figure*}

\begin{figure*}[t!]
\centering
\includegraphics[width=1.15\textwidth,trim={70 0 0 0},clip]{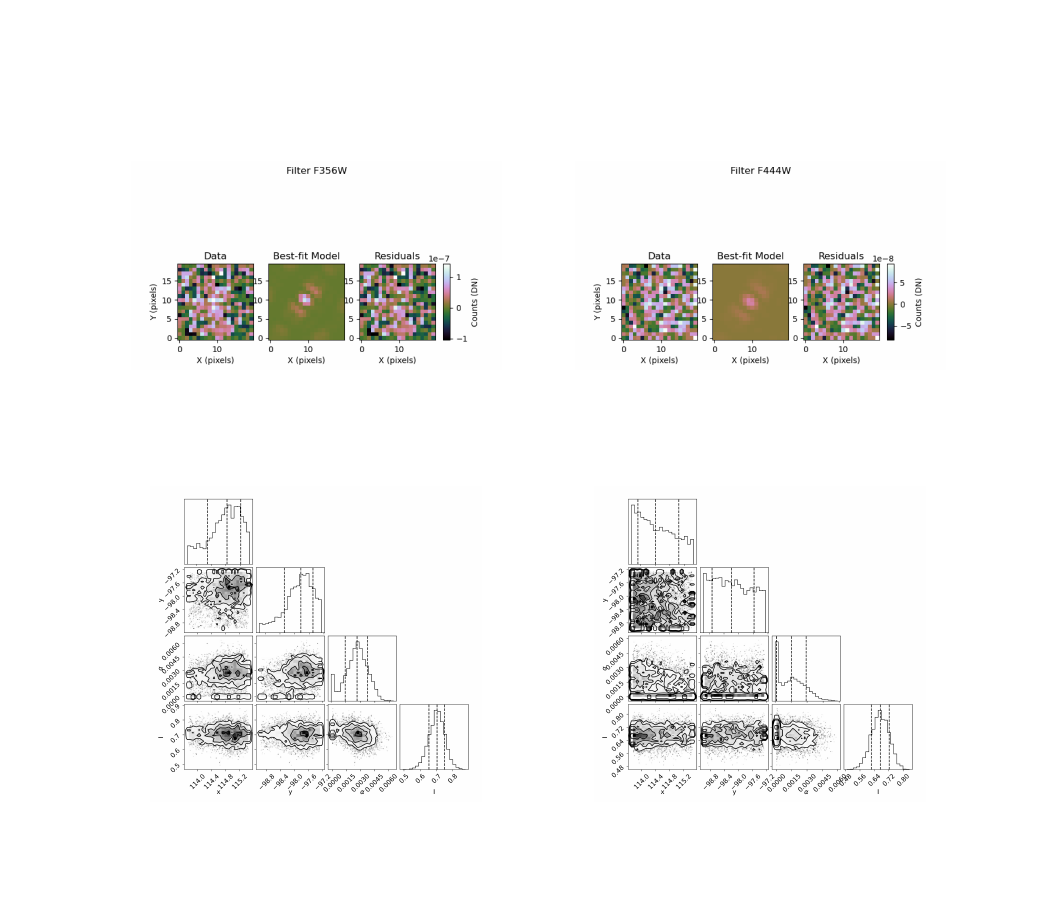}
\vskip -0.4in
\caption{Same as Figure~\ref{fig:mcmc1}, but for \textit{S2}.}
\label{fig:mcmc2}
\end{figure*}

\begin{figure*}[t!]
\centering
\includegraphics[width=1.15\textwidth,trim={70 0 0 0},clip]{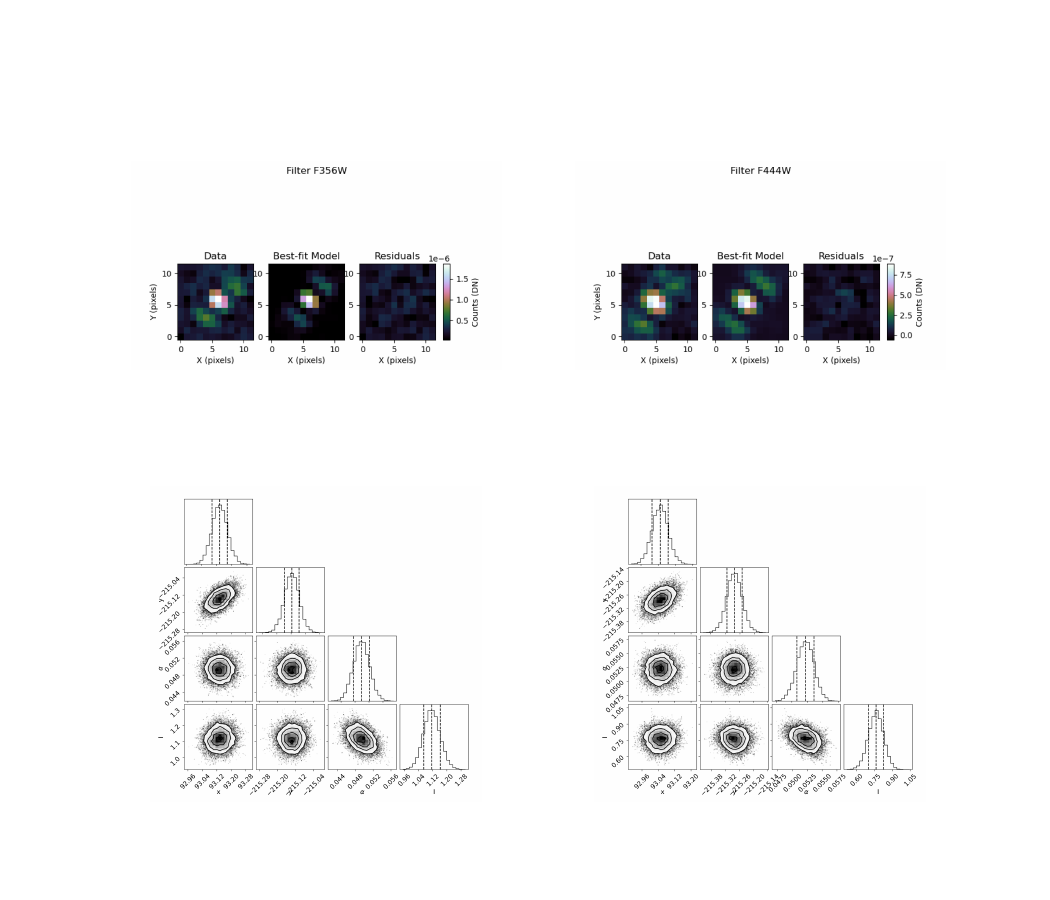}
\vskip -0.4in
\caption{ 
Same as Figure~\ref{fig:mcmc1}, but for source HST \#1.
\label{fig:mcmchst1}}
\end{figure*}

Figures~\ref{fig:mcmc1},~\ref{fig:mcmc2},~~and~\ref{fig:mcmchst1} show the output of \texttt{pyKLIP}'s astrometry and photometry model fit. The comparison of the PSF-subtracted image versus the forward model, and its residuals, give a visual representation of the the source extraction accuracy. The corner plots show how well the MCMC walkers converge to a solution. For  \textit{S1} and \textit{S8} for instance, the MCMC iterations converge well given the SNR of these detection. 

\begin{figure*}[t!]
\centering
\includegraphics[width=1.15\textwidth,trim={70 0 0 0},clip]{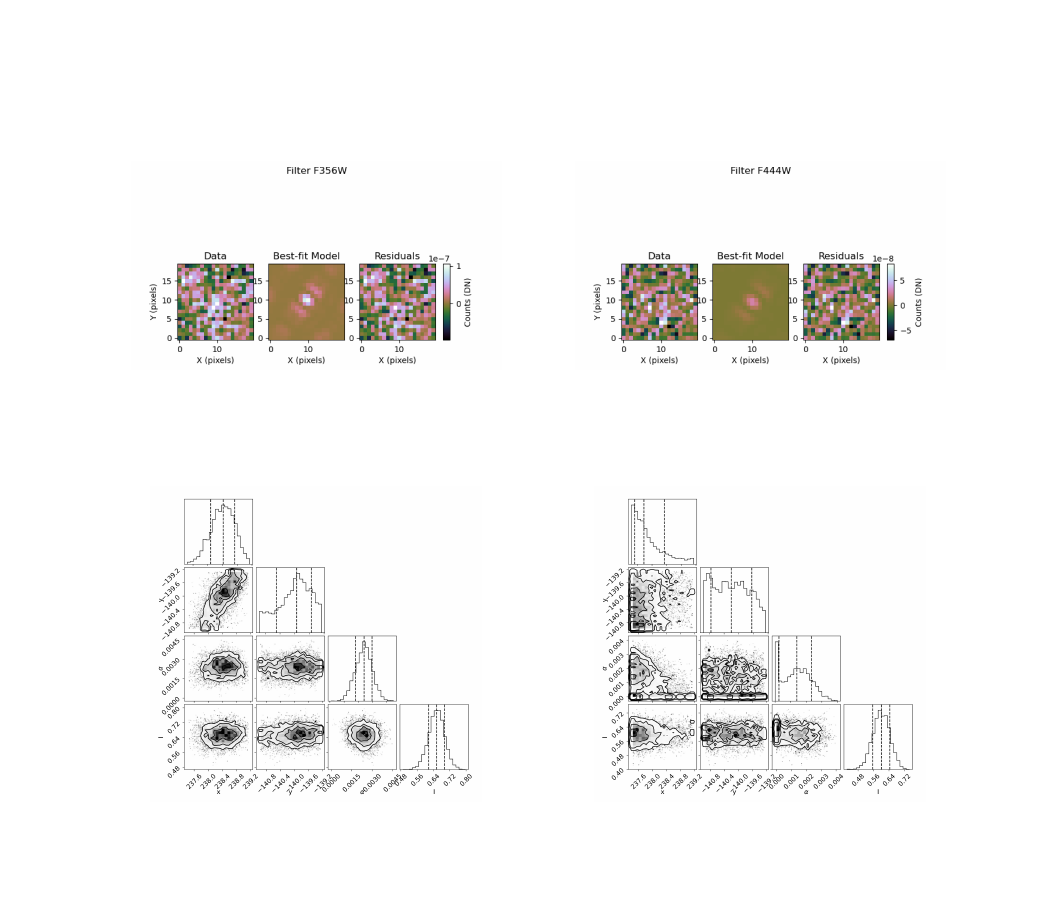}
\vskip -0.4in
\caption{ 
Same as Figure~\ref{fig:mcmc1}, but for source 3.
\label{fig:mcmc3}}
\end{figure*}

\begin{figure*}[t!]
\centering
\includegraphics[width=1.15\textwidth, trim={70 0 0 0}, clip]{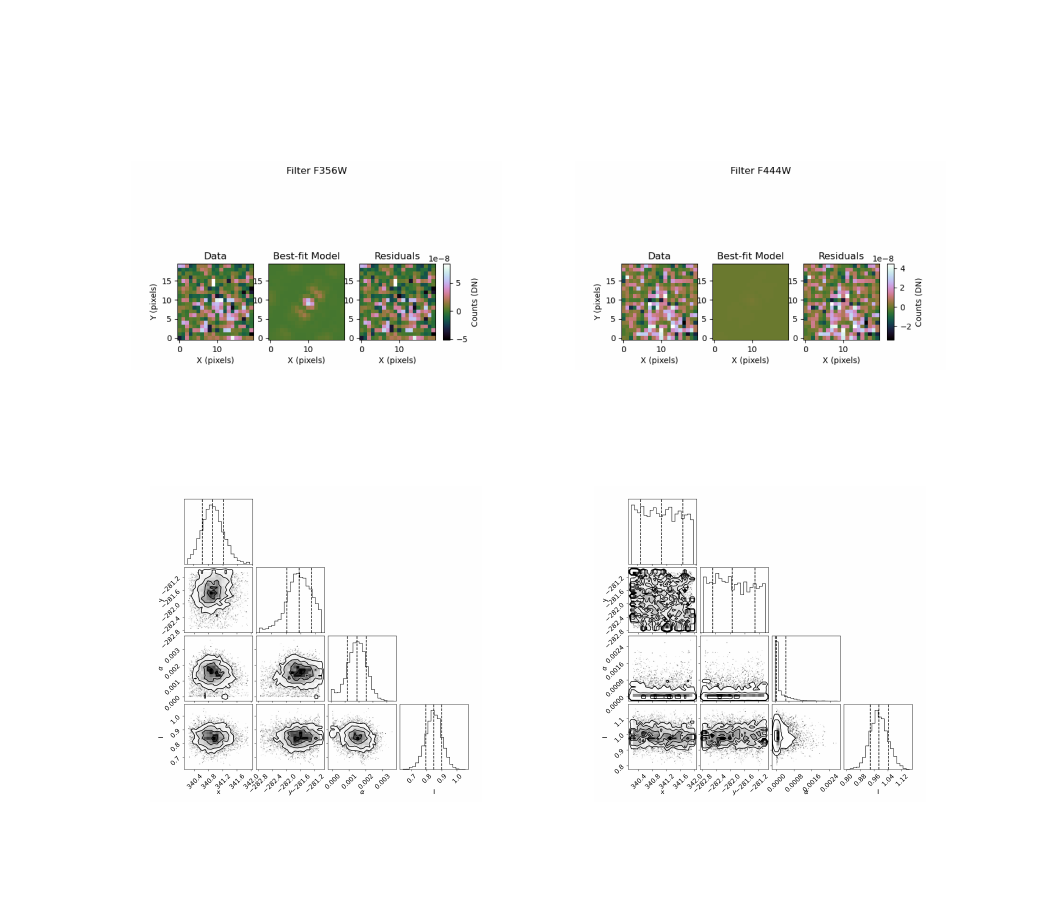}
\vskip -0.4in
\caption{ 
Same as Figure~\ref{fig:mcmc1}, but for source 4.
\label{fig:mcmc4}}
\end{figure*}

\begin{figure*}[t!]
\centering
\includegraphics[width=1.15\textwidth,trim={70 0 0 0},clip]{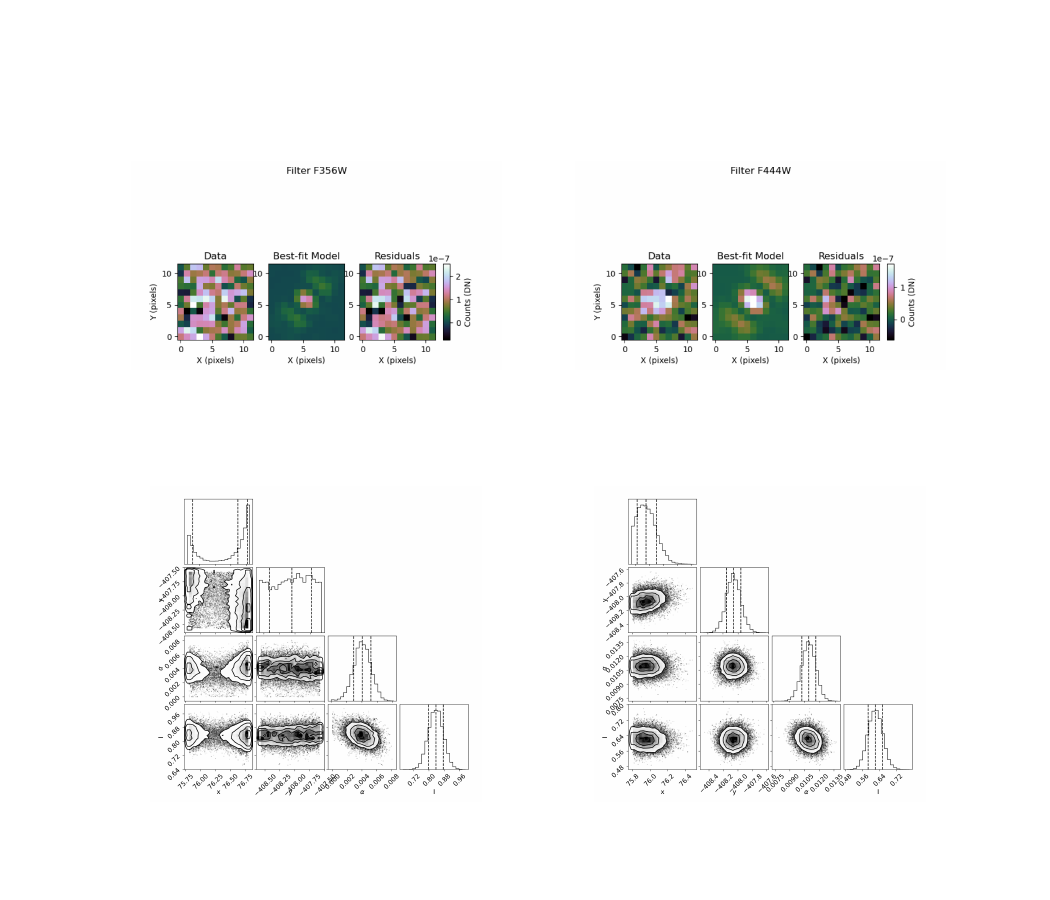}
\vskip -0.4in
\caption{ 
Same as Figure~\ref{fig:mcmc1}, but for source 5.
\label{fig:mcmc5}}
\end{figure*}

For completeness, we include the results of the model fits for sources 3, 4, and 5, in Figures~\ref{fig:mcmc3},~\ref{fig:mcmc4},~\ref{fig:mcmc5}. Although the photometry and astrometry of these sources was ultimately recovered with an aperture photometry based method, we report the MCMC model fit result figures to showcase the spurious uncertainties that stem from the low flux combined with the probable extended nature of these sources.


\end{document}